\documentclass[a4paper,11pt]{article}
\pdfoutput=1 % if your are submitting a pdflatex (i.e. if you have
\usepackage{jheppub} % for details on the use of the package, please
\usepackage[T1]{fontenc} % if needed

\usepackage{amsthm} %for theorems and demonstrations 
\usepackage{amsmath}

\newtheorem{mytheorem}{Property}

\newcommand{\g}{g}

\newcommand{\ip}{\Gamma}
\newcommand{\ai}{a_{\infty}}

\newcommand\M{\multicolumn{1}{c}}
\newcommand\MI{\multicolumn{1}{c|}}

\usepackage{tikz}
\usetikzlibrary{math}
\usetikzlibrary{arrows,shapes,positioning}
\usetikzlibrary{decorations.markings}
\tikzstyle arrowstyle=[scale=1]
\tikzstyle directed=[postaction={decorate,decoration={markings,
    mark=at position .65 with {\arrow[arrowstyle]{stealth}}}}]
\tikzstyle reverse directed=[postaction={decorate,decoration={markings,
    mark=at position .65 with {\arrowreversed[arrowstyle]{stealth};}}}]

\newlength{\mywidth}
\usepackage[colorlinks=true
,urlcolor=blue
,anchorcolor=blue
,citecolor=blue
,filecolor=blue
,linkcolor=blue
,menucolor=blue
,pagecolor=blue
,linktocpage=true
,pdfproducer=medialab
,pdfa=true
]{hyperref}

\title{\boldmath One-loop integrability with shifting masses}

\author[a,b]{Matheus Fabri,}
\author[a,b,c]{Davide Polvara}

\affiliation[a]{Dipartimento di Fisica e Astronomia,
Universita degli Studi di Padova, via Marzolo 8, 35131 Padova, Italy.}
\affiliation[b]{INFN,
Sezione di Padova, via Marzolo 8, 35131 Padova, Italy.}
\affiliation[c]{II. Institut f\"ur Theoretische Physik,  Universit\"at Hamburg, Luruper Chaussee 149, 22761 Hamburg, Germany.}

\emailAdd{matheusaugusto.fabri@unipd.it}
\emailAdd{davide.polvara@gmail.com}

\abstract{We investigate the perturbative integrability of two-dimensional massive quantum field theories with polynomial-like interactions and show that any theory of such class which is purely elastic at the tree level is also purely elastic at one loop.
To preserve the elasticity, the physical renormalized masses of the theory must differ from the classical ones by quantum corrections carried by one-loop bubble diagrams. After the masses are corrected in this manner we show that one-loop inelastic processes vanish and integrability is preserved under one-loop effects. 
Relying on this fact we show that the closed expression for one-loop S-matrices in terms of tree S-matrices obtained in~\cite{Fabri:2024qgd} extends to models that do not preserve the mass ratios at one loop. We test our results on the full class of nonsimply-laced affine Toda theories and find exact match with the S-matrices bootstrapped in the past.}

\begin{document} \begin{flushright}\small{ZMP-HH/24-27}\end{flushright}
\maketitle
\flushbottom

\section{Introduction}

Integrable models in 1+1 dimensions are among the few examples of quantum field theories which can be solved exactly.
They are characterised by an infinite tower of higher-spin conserved charges in involution.
If these charges survive the quantization, their combination with the additional requirements of unitarity, crossing and analyticity can be used to find the exact two-body S-matrix of these theories through the \textit{bootstrap program}~\cite{Zamolodchikov:1978xm}. Multibody S-matrices factorize then into the product of two-body S-matrices~\cite{Parke:1980ki} and the scattering is completely solved. While the bootstrap philosophy has been useful for the determination of the exact S-matrices of many integrable theories, from the sine-Gordon model~\cite{Zamolodchikov:1978xm} to the more exotic theories living on the worldsheet of superstrings~\cite{Beisert:2010jr}, 
the question on whether a generic classically integrable theory is also quantum integrable has never been completely answered so far. 
More specifically, the question we ask ourselves is the following: `For a given theory with purely elastic scattering at the tree level, can we be sure that the scattering is purely elastic also at one loop?'.
This question was recently answered only for massive bosonic models with polynomial-like interactions with mass ratios unaffected by one-loop corrections~\cite{Polvara:2023vnx,Fabri:2024qgd}. In this paper, we extend the answer to theories with mass ratios that receive arbitrary nontrivial quantum corrections and find that the answer to the above question is still `yes'.
After having established that any tree-level integrable theory of this type is also one-loop integrable we provide a closed expression for its one-loop S-matrix as a function of its tree-level S-matrix, generalising the results of~\cite{Fabri:2024qgd} to models receiving non-trivial corrections to the mass ratios. We test our formulas on the class of non-simply-laced affine Toda theories, finding exact agreement with the results bootstrapped in the past~\cite{Delius:1991kt,Corrigan:1993xh} for this class of models. 

The paper is structured as follows: in section~\ref{OneLIntConstraints} we present the class of tree-level integrable models we consider and show that all of them are also one-loop integrable provided the renormalized masses receive quantum corrections from bubble diagrams. After establishing the renormalized masses of these models, we write a closed expression for their one-loop S-matrix in terms of tree-level data. We show that Landau singularities in one-loop inelastic amplitudes are completely encoded in the mass corrections and vanish in the total amplitude only if we set the values of the physical masses to depend on the coupling. In section~\ref{OneloopS_mat_nonsimplaced} we test our formula on the class on non-simply-laced affine Toda theories, finding agreement with the known results in the literature~\cite{Delius:1991kt,Corrigan:1993xh}. We conclude in section~\ref{sec:Conclusions}, presenting the main results and open problems. In appendix~\ref{app:Jacobian} we review the simple derivation of the Jacobians appearing in S-matrices of 1+1 dimensional theories.
In appendices~\ref{app:geometrical_relations} and~\ref{app:mass-shifts} we review the notion of tree-level integrability and provide a complete list of the mass corrections of nonsimply-laced affine Toda theories, which we extract from the tree-level S-matrix in the way proposed in~\cite{Polvara:2023vnx}.

\section{One-loop integrability constraints}
\label{OneLIntConstraints}

We start from a bare Lagrangian of the following type
\begin{equation}
\label{eq0_1}
\mathcal{L}= \sum^r_{a=1} \left(\frac{\partial_\mu \phi_a  \partial^\mu \phi_{\bar{a}}}{2} - \frac{m_a^2}{2}  \phi_a \phi_{\bar{a}}  \right) - \sum_{n=3}^{+\infty} \frac{1}{n!}\sum_{a_1,\dots, a_n=1}^r C^{(n)}_{a_1 \ldots a_n} \phi_{a_1} \ldots \phi_{a_n} \,,
\end{equation}
where $a=1, \, \dots,\, r$ is a label for the particle types, which correspond to the asymptotical states of the theory. For these labels we follow the same notation of~\cite{Dorey:2021hub}, which is $a=\bar{a} \in \{1, \, \dots, \, r \}$ if $\phi_a$ is real and $a\ne\bar{a}$ (i.e. $a$ and $\bar{a}$ are different indices both contained in the set $\{1, \dots, r\}$) if $\phi_a$ is complex. In this second case we define $\phi_{\bar{a}}=\phi^*_a$. 
The couplings are chosen to satisfy $C_{\bar{a}
_1  \dots \bar{a}_n}^{(n)}=(C_{a_1 \dots a_n}^{(n)})^*$ so to guarantee the reality of the Lagrangian. To be more specific consider the example in which there are three particles in the spectrum ($1$, $2$ and $3$) such that $\phi_3$ is a real field (for which $\phi_3=\phi_{\bar{3}}$) and the fields $\phi_1$ and $\phi_2$ are one the complex conjugate of the other and satisfy $\phi_2=\phi_{\bar{1}}=\phi^*_1$ and $\phi_1=\phi_{\bar{2}}=\phi^*_2$. In this case the kinetic term takes the following form
\begin{equation}
\begin{split}
&\left(\frac{\partial_\mu \phi_1  \partial^\mu \phi_{\bar{1}}}{2} - \frac{m_1^2}{2}  \phi_1 \phi_{\bar{1}}  \right)+\left(\frac{\partial_\mu \phi_2  \partial^\mu \phi_{\bar{2}}}{2} - \frac{m_2^2}{2}  \phi_2 \phi_{\bar{2}}  \right)+\left(\frac{\partial_\mu \phi_3  \partial^\mu \phi_{\bar{3}}}{2} - \frac{m_3^2}{2}  \phi_3 \phi_{\bar{3}}\right)\\
&=\left(\partial_\mu \phi_1  \partial^\mu \phi^*_{1} - m_1^2  \phi_1 \phi^*_{1}  \right)+\left(\frac{\partial_\mu \phi_3  \partial^\mu \phi_{3}}{2} - \frac{m_3^2}{2}  \phi_3 \phi_{3}\right)\,,
\end{split}
\end{equation}
which is the standard real Lagrangian for a complex and a real field. Similarly it is easy to show that the interacting part is real too and the theory is unitary.

The masses and couplings are set so that the theory is purely elastic at the tree level, by which we mean that the only tree-level amplitudes different from zero are those in which the incoming and outgoing particles are of the same type and carry the same set of momenta. This requires the masses of the theory to satisfy the so-called \textit{flipping rule}~\cite{Braden:1990wx}, necessary for having simultaneous poles in different Feynman diagrams contributing to inelastic tree-level amplitudes, as briefly reviewed in appendix~\ref{app:tree_canc_and_flipping}. The condition for the residues of these poles to cancel constrains the $3$-point couplings of the theory, which can be defined in terms of the masses~\cite{Dorey:2021hub}. Higher-order couplings are then obtained inductively by requiring the vanishing of all tree production amplitudes with multiple external legs~\cite{Dorey:1996gd,Bercini:2018ysh, Gabai:2018tmm}.

The space of Lagrangians of type~\eqref{eq0_1} satisfying the requirement of purely tree-level elasticity is non-empty. For example, all the affine Toda theories are purely elastic at the tree-level, as proven in~\cite{Dorey:2021hub}. Minimal S-matrices of Toda type appeared in the past to describe integrable deformations of minimal models (see, e.g. \cite{Zamolodchikov:1989fp} and \cite{Christe:1989ah})\footnote{Strictly speaking, the S-matrices describing minimal models deformations and the ones of affine Toda type agree on the poles and mass spectrum, but not on the zeros. The latter are indeed deformations of minimal S-matrices obtained by adding extra zeros in physical strip~\cite{Braden:1989bu}.}. 
These theories are among the simplest examples of quantum integrable models and their perturbative understanding would provide a starting point to approach more complicated theories, such as sigma models in string theory (for which derivative interactions and fermionic fields must be included).

In this section, we show that any theory with a Lagrangian of type~\eqref{eq0_1} which is purely elastic at the tree level is also purely elastic at one loop.
Relying on this fact we provide a closed expression for the one-loop S-matrix of this class of theories in terms of tree-level data, extending the results of~\cite{Fabri:2024qgd} to models having masses which do not scale equally under one-loop corrections.

\subsection{Renormalized perturbation theory and one-loop inelastic processes}

We study one-loop amplitudes working with renormalized perturbation theory.   
We follow the renormalization conditions discussed in~\cite{Fabri:2024qgd}, which we summarize below:
\begin{itemize}
    \item[(1)] All amplitudes must be free of UV divergences at one-loop;
    %%%%%%%%%%
    \item[(2)] $\forall \ a, b \in \{1, \dots, r\}$ the propagator $G_{ab}(p^2)=G^{(0)}_{ab}(p^2)+G^{(1)}_{ab}(p^2)$, composed of a tree-level and one-loop contribution, must have
    \begin{equation}
    \label{eq:res_prop_rencond}
\text{Res} \, G_{ab}(p^2)\Bigl|_{p^2=\hat{m}^2_a}=\text{Res} \, G_{ab}(p^2)\Bigl|_{p^2=\hat{m}^2_b}= i \delta_{ab} \,,
    \end{equation}
    where $\hat{m}_a$ and $\hat{m}_b$ are the renormalised masses;
    %%%%%%%%%%
    \item[(3)] all inelastic amplitudes need to vanish at one loop for all values of momenta satisfying the on-shell condition $p_j^2=\hat{m}^2_j$.
\end{itemize}
The first condition can always be satisfied for a proper choice of counterterms cancelling all diagrams with self-contracted vertices; this condition can be equally implemented by introducing a normal ordering in the Lagrangian interaction. Due to this fact, we omit these diagrams and neglect to write the UV counterterms explicitly.  A more detailed discussion about these counterterms can be found for example in~\cite{Polvara:2023vnx, Christe:1989my}.
The second and third conditions above require to write the bare masses and couplings appearing in the Lagrangian~\eqref{eq0_1} as
\begin{equation}
\label{eq:ren_masses_couplings}
\begin{split}
&m_a^2=\hat{m}^2_a+ \delta m^2_a\,,\\
&C^{(n)}_{a_1 \dots a_n}=\hat{C}^{(n)}_{a_1 \dots a_n}+ \delta C^{(n)}_{a_1 \dots a_n} \,,
\end{split}
\end{equation}
where $\hat{m}^2_a$ and $\hat{C}^{(n)}_{a_1 \dots a_n}$ are the renormalized masses and couplings. We fix the renormalized masses and couplings, and associated counterterms, in such a way as to prevent inelastic processes at the tree level and one loop.

We consider inelastic processes of the following form
\begin{equation}
\label{eq:ab_to_cd_process}
a(p)+b(p') \rightarrow c(q)+d(q')\,,
\end{equation}
with $\{a,b\} \ne \{c, d\}$. 
Following the notation of~\cite{Fabri:2024qgd}, the one-loop renormalized amplitude associated with the process~\eqref{eq:ab_to_cd_process} can be written as 
\begin{equation}
\label{eq:full_one_loop_amplitude}
    M^{(1)}_{ab\rightarrow cd} = M_{ab\rightarrow cd}^{\textrm{(1-loop)}} + M_{ab\rightarrow cd}^{\textrm{(ct.I)}} + M_{ab\rightarrow cd}^{\textrm{(ct.II)}},
\end{equation}
where the first term on the RHS of the expression above comes from summing over all connected one-loop diagrams obtained by applying standard Feynman rules on the Lagrangian~\eqref{eq0_1}. The term $ M_{ab\rightarrow cd}^{\textrm{(ct.I)}}$ contains instead diagrams with two-point counterterms.  Finally, the term $M_{ab\rightarrow cd}^{\textrm{(ct.II)}}$ contains tree-level diagrams with coupling counterterms (for two-to-two amplitudes these counterterms only include $\delta C^{(3)}_{a_1 a_2 a_3}$ and $\delta C^{(4)}_{a_1  a_2  a_3 a_4}$).
While $M_{ab\rightarrow cd}^{\textrm{(1-loop)}}$ and $M_{ab\rightarrow cd}^{\textrm{(ct.I)}}$ are separately ill-defined (since both of them contain divergent contributions), their sum is finite; in all theories with Lagrangians of type~\eqref{eq0_1} which are purely elastic at the tree level this sum is given by~\cite{Polvara:2023vnx}
\begin{equation}
\label{ab_to_cd_one_loop_inelastic_amp_derivatives}
    M_{ab\rightarrow cd}^{\textrm{(1-loop)}} + M_{ab\rightarrow cd}^{\textrm{(ct.I)}} =  \sum_{k \in \{\text{prop}, \text{ext}\}} \delta m_k^2 \frac{\partial}{\partial \mu^2_k} M_{ab \to cd}^{(0)}\Bigl|_{\mu^2_k=\hat{m}^2_k},
\end{equation}
with the summation done over the masses appearing in propagators ($k\in\textrm{prop}$) and external states ($k\in\textrm{ext}$). The latter appears in the amplitude through the Mandelstam variables. Here $\mu_k$ are free parameters which are set to the renormalized masses $\hat{m}_k$ only after performing the derivatives. 
Note that the RHS of the expression above can be evaluated either at the values of the renormalized masses $\hat{m}_k$ or at the values of the classical masses ${m}_k$. This is irrelevant since the difference only appears from the two-loop order.
The set $\{\delta m_a^2 \}^r_{a=1}$ is composed of the radiative corrections to the masses, computed by summing over one-loop bubble diagrams.
 In 1+1 dimensions, it is convenient to parameterise the momentum $p=(p_0, p_1)$ of a certain particle of type $j$ by
\begin{equation}
p_0= \mu_j \cosh{\theta_p} \,, \qquad p_1=\mu_j \sinh{\theta_p}
\end{equation}
so that the following condition holds
\begin{equation}
p^2_0-p^2_1=\mu_j^2 \,.
\end{equation}
We refer to the parameter $\mu_j$ as the off-shell mass of the particle. For $\mu_j = m_j$ the particle becomes on-shell. 

We define the Mandelstam variables $s$, $t$, and $u$ using the off-shell mass values $\mu_k$ as
\begin{align}
\label{eq:s_t_u_Mandelstam}
    s &  = (p+p')^2= \mu_a^2 + \mu_b^2 + 2 \mu_a \mu_b \cosh\theta_{pp'}, \\
    t &  = (p-q')^2= \mu_a^2 + \mu_d^2 - 2 \mu_a \mu_d \cosh\theta_{pq'}, \\
    u & = (p-q)^2=\mu_a^2 + \mu_c^2 - 2 \mu_a \mu_c \cosh\theta_{pq}.
\end{align}
The parameters $\theta_p$, $\theta_{p'}$, $\theta_q$ and $\theta_{q'}$ are the rapidities of the scattered particles (whose momenta are $p$, $p'$, $q$ and $q'$ respectively).
Moreover we define $\theta_{p p'} \equiv \theta_p - \theta_{p'}$. Note that two of the four rapidities $\{ \theta_{p} , \theta_{p'} , \theta_{q} , \theta_{q'} \}$ are fixed by momentum conservation. We will fix the outgoing rapidities as functions of the incoming ones, which means that $\theta_{pp'}$ is free and $\{\theta_{pq'}, \theta_{pq} \}$ depend on $\theta_{pp'}$ and the external masses through momentum conservation. 

Formula~\eqref{ab_to_cd_one_loop_inelastic_amp_derivatives} was originally obtained in~\cite{Polvara:2023vnx}, thanks to the splitting of each Feynman propagator into a retarded propagator and a Dirac delta function:
\begin{equation}
\frac{i}{p^2 - m^2 + i\epsilon}= \frac{i}{p^2 - m^2 - i p_0 \epsilon}+ \frac{\pi}{\sqrt{p_1^2+m^2}} \delta(p_0 - \sqrt{p_1^2+m^2})\,.
\end{equation}
It easy to show that diagrams containing only propagators of retarded type have poles in the same half of the energy complex plane and it is possible to compute their integrals by closing a contour containing no poles. Due to this fact, diagrams of this type are zero and the one-loop amplitude can be written as
\begin{equation}
\label{eq:splitting_ono_loop_amplitude_into_double_and_single_cuts}
\begin{split}
M_{ab\rightarrow cd}^{\textrm{(1-loop)}}&=
\sum_{e,f=1}^r \frac{1}{8 m_e m_f |\sinh{\theta_{ef}}|} \left(M_{ab\to ef}^{(0)}M_{ef\to cd}^{(0)}  +M_{ae\to cf}^{(0)}M_{bf\to de}^{(0)} + M_{ae\to df}^{(0)}M_{bf\to ce}^{(0)}\right)\\
&+\frac{1}{8 \pi} \sum^r_{e=1} \int^{+\infty}_{-\infty} d \theta_k M_{ab e \to cd e}^{(0)}(p , p' , k; q, q', k) \,.
\end{split}
\end{equation}
The first line of the expression above is associated with contributions containing two Dirac delta functions while the second line comes from contributions containing a single Dirac delta function, for which we still need to integrate over a one-dimensional space. Since the theory is purely elastic at the tree-level and $\{a, b\} \ne \{c, d\}$ then the first line in~\eqref{eq:splitting_ono_loop_amplitude_into_double_and_single_cuts} vanishes (indeed each term in parenthesis contains at least one inelastic two-to-two amplitude which is zero by our starting assumption). One may expect that this is the case also for the integrand in the second line of~\eqref{eq:splitting_ono_loop_amplitude_into_double_and_single_cuts} (note indeed that the tree-level amplitude $M_{ab e \to cd e}^{(0)}(p , p' , k; q, q', k)$ is inelastic). However, certain tree-level Feynman diagrams composing the integrand are ill-defined since they contain collinear singularities (an example is reported in figure~\ref{fig:collinear_singularity_3_to_3_tree_level}) and it is not possible to claim that the second line of~\eqref{eq:splitting_ono_loop_amplitude_into_double_and_single_cuts} is zero. After removing these ill-defined terms thanks to the counterterms in $M_{ab\rightarrow cd}^{\textrm{(ct.I)}}$ and using the tree-level integrability of the theory to compute the well-defined part of $M_{ab e \to cd e}^{(0)}$ one ends up with formula~\eqref{ab_to_cd_one_loop_inelastic_amp_derivatives}. 
\begin{figure}
\begin{center}
\begin{tikzpicture}
\tikzmath{\y=1.5;}

%Diagram 1
\draw[directed] (-1*\y,1*\y) -- (0*\y,0*\y);
\draw[directed] (0*\y,0*\y) -- (0*\y,1*\y);
\draw[directed] (0*\y,-1*\y) -- (0*\y,0*\y);
\draw[directed] (0*\y,0*\y) -- (1*\y,-1*\y);
\draw[directed] (0*\y,-2*\y) -- (1*\y,-1*\y);
\draw[directed] (1*\y,-1*\y) -- (3*\y,-1*\y);
\draw[directed] (3*\y,-1*\y) -- (4*\y,0*\y);
\draw[directed] (3*\y,-1*\y) -- (4*\y,-2*\y);
\filldraw[black] (-1.5*\y,1*\y)  node[anchor=west] {\scriptsize{$a(p)$}};
\filldraw[black] (-0.5*\y,1*\y)  node[anchor=west] {\scriptsize{$e(k)$}};
\filldraw[black] (-0.5*\y,-1*\y)  node[anchor=west] {\scriptsize{$e(k)$}};
\filldraw[black] (0.5*\y,-0.5*\y)  node[anchor=west] {\scriptsize{$a$}};
\filldraw[black] (-0.4*\y,-1.8*\y)  node[anchor=west] {\scriptsize{$b(p')$}};
\filldraw[black] (4*\y,-1.8*\y)  node[anchor=west] {\scriptsize{$c(q)$}};
\filldraw[black] (4*\y,0.2*\y)  node[anchor=west] {\scriptsize{$d(q')$}};

\end{tikzpicture}
\caption{Example of collinear singularity in the integrand of one-loop amplitudes. The momentum of the particle $a$ propagating internally to the diagram is equal to $p$ and is the same as the external momentum. Then the internal propagating particle $a$ is on-shell and the diagram is singular.}
\label{fig:collinear_singularity_3_to_3_tree_level}
\end{center}
\end{figure}
We remark that this result is not obtained by unitarity cut methods; instead all contributions of Feynman diagrams are
taken into account and formula~\eqref{eq:splitting_ono_loop_amplitude_into_double_and_single_cuts} carries complete information on the amplitude (both its rational and irrational part).

Adding to the quantity in~\eqref{ab_to_cd_one_loop_inelastic_amp_derivatives} the tree-level amplitude\footnote{Note that the tree-level amplitude must be evaluated on-shell at the values of the renormalized masses and renormalized couplings.}
\begin{equation}
\label{eq:tree_lev_eval_at_ren}
M^{(0)}_{ab \to cd}\Bigl|_{\hat{m}^2_k, \hat{C}^{(n)}}
\end{equation}
and all diagrams with coupling counterterms, which we collect in
\begin{multline}
\label{eq:int-ct_new}
    M_{ab\rightarrow cd}^{\textrm{(ct.II)}} = -i \sum_{i\in s} \frac{\delta C^{(3)}_{ab\bar{i}} \hat{C}^{(3)}_{i\bar{c}\bar{d}} + \hat{C}^{(3)}_{ab\bar{i}} \delta C^{(3)}_{i\bar{c}\bar{d}}}{s-m_i^2} 
    -i \sum_{j\in t} \frac{\delta C^{(3)}_{a\bar{d}\bar{j}} \hat{C}^{(3)}_{jb\bar{c}} + \hat{C}^{(3)}_{a\bar{d}\bar{j}} \delta C^{(3)}_{jb\bar{c}}}{t-m_j^2} \\
    -i \sum_{l\in u} \frac{\delta C^{(3)}_{a\bar{c}\bar{l}} \hat{C}^{(3)}_{lb\bar{d}} + \hat{C}^{(3)}_{a\bar{c}\bar{l}} \delta C^{(3)}_{lb\bar{d}}}{u-m_l^2} - i \delta C^{(4)}_{ab\bar{c}\bar{d}}\,,
\end{multline}
we obtain\footnote{ This comes from noting that the sum of~\eqref{ab_to_cd_one_loop_inelastic_amp_derivatives} and~\eqref{eq:int-ct_new} corresponds to the next-to-leading order expansion of $M_{ab \to cd}^{(0)}$ around the renormalized values of the masses and couplings.}
\begin{equation}
\label{ab_to_cd_one_loop_inelastic_amp_without_derivatives}
M_{ab \to cd}^{(0)}\Bigl|_{\hat{m}^2_k, \hat{C}^{(n)}}+ M_{ab \to cd}^{(1)} = M_{ab \to cd}^{(0)}\Bigl|_{\hat{m}^2_k+\delta m^2_k,\, \hat{C}^{(n)}+\delta C^{(n)}}+ \dots \,.
\end{equation}
The ellipses in the expression above contain terms starting contributing at the two-loop order. Note that the tree amplitude on the RHS of~\eqref{ab_to_cd_one_loop_inelastic_amp_without_derivatives} is evaluated at \emph{shifted} values of the physical masses $\hat{m}_k^2$ and physical couplings $\hat{C}^{(n)}$.
Recalling that the classical masses and couplings are connected to the renormalized
ones through the relation~\eqref{eq:ren_masses_couplings} we obtain that the inelastic amplitude to one-loop order in
perturbation theory is given by
\begin{equation}
\label{ab_to_cd_one_loop_inelastic_amp_without_derivatives2}
\begin{split}
M_{ab \to cd}^{(0)}\Bigl|_{\hat{m}^2_k, \hat{C}^{(n)}}+ M_{ab \to cd}^{(1)} &=M_{ab \to cd}^{(0)}\Bigl|_{m^2_k,\, C^{(n)}}+ \mathcal{O}((\delta m^2_k)^2)\,.
\end{split}
\end{equation}
The quantity on the RHS of the expression above is a tree-level amplitude evaluated
at the classical values of the masses and couplings appearing in~\eqref{eq0_1} and vanishes by our starting assumption that the theory is purely elastic at the tree level.
Due to this fact, the model is purely elastic at one loop by construction.
We conclude that \textit{any theory with a Lagrangian of type~\eqref{eq0_1} which is purely elastic at
the tree level is also purely elastic at one loop}.

In the discussion above, we focused on two-to-two one-loop inelastic amplitudes; however, our argument smoothly extends to one-loop production amplitudes with an arbitrary number of external legs. The basic ingredient for proving the absence of inelasticity is indeed the result of~\cite{Polvara:2023vnx}, which is equally valid for production amplitudes. With our conventions, equation (1.6) of that paper is given by
\begin{equation}
\label{ab_to_cd_one_loop_inelastic_amp_derivatives_prod}
    M_{a_1 a_2\rightarrow a_3 \dots a_n}^{\textrm{(1-loop)}} + M_{a_1 a_2\rightarrow a_3 \dots a_n}^{\textrm{(ct.I)}} =  \sum_{k \in \{\text{prop}, \text{ext}\}} \delta m_k^2 \frac{\partial}{\partial \mu^2_k} M_{a_1 a_2\rightarrow a_3 \dots a_n}^{(0)}\Bigl|_{\mu^2_k=\hat{m}^2_k},
\end{equation}
and the previous discussion extends to one-loop production processes.

We stress that the classical masses are coupling independent and therefore the renormalized masses $\{\hat{m}_a \}^r_{a=1}$ (defined through equation~\eqref{eq:ren_masses_couplings}) depend on the coupling through the radiative corrections $\{\delta m_a\}^r_{a=1}$.
It is quite remarkable that the
final amplitude on the RHS of~\eqref{ab_to_cd_one_loop_inelastic_amp_without_derivatives2} vanishes no matter the mass- and coupling-corrections
since these corrections disappear from the final result.

For this to be the case, it was fundamental that the one-loop amplitude $M_{ab \to cd}^{(1)}$ takes the specific form obtained by summing~\eqref{ab_to_cd_one_loop_inelastic_amp_derivatives} and~\eqref{eq:int-ct_new}, and corresponds in this way to the variation of a tree-level amplitude around the renormalized values of the masses and couplings. This is a nontrivial result, emerging after summing over all one-loop Feynman diagrams and requiring the amplitudes to be purely elastic at the tree level. 
One may expect that a necessary condition for a theory to remain integrable at loop level is that the mass ratios do not renormalize so that the S-matrix poles take fixed values in the physical strip not depending on the coupling. From our result, we see that this is not the case. Instead, the classical masses can receive arbitrary one-loop quantum corrections without spoiling the integrability of the model. Examples of exact S-matrices with physical poles being coupling dependent are known from the past~\cite{Delius:1991kt}; however, a complete check of the absence of inelastic processes was never achieved. Moreover, it was unclear so far under which conditions the mass corrections $\{\delta m^2_k\}^r_{k=1}$ (obtained by summing over one-loop bubble diagrams) are compatible with the absence of inelasticity in one-loop processes. Our results indicate that no condition on $\{\delta m^2_k\}^r_{k=1}$ is required.

Let us also stress that the result on the LHS of~\eqref{ab_to_cd_one_loop_inelastic_amp_without_derivatives2} contains the sum of a tree-level amplitude ($M_{ab \to cd}^{(0)}$) evaluated at the renormalized values of the masses and couplings and a one-loop amplitude ($M_{ab \to cd}^{(1)}$). As it is well known $M_{ab \to cd}^{(0)}$ contains only rational terms in the Mandelstam variables, which appear as hyperbolic functions of the rapidities (i.e. rational functions of $\cosh{\theta}$). On the opposite $M_{ab \to cd}^{(1)}$ can also contain branch cuts, which normally become rational functions of the rapidities (i.e. rational functions of $\theta$). The only possibility for the sum of $M_{ab \to cd}^{(0)}$ and $M_{ab \to cd}^{(1)}$ to cancel is therefore that $M_{ab \to cd}^{(1)}$ is free of any rational term of $\theta$. From our results we see that this is always the case as far as at tree-level the theory is purely elastic. This can also be justified by standard unitarity cut methods; indeed the two-to-two tree-level amplitudes arising by cutting one-loop inelastic amplitudes are also inelastic and therefore zero by our starting assumption. Then unitarity predicts the absence of any branch cut in theories with purely-elastic tree-level S-matrices.

\subsection{One-loop elastic amplitudes}

Now we discuss one-loop elastic S-matrices for processes of the following type
\begin{equation}
a(p)+b(p') \to a(p)+ b(p')\,.
\end{equation}
We stress that the S-matrix is connected to the amplitude through\footnote{For elastic amplitudes and S-matrices we use the notation $M_{ab} \equiv M_{ab \to ab}$ and $S_{ab} \equiv S_{ab \to ab}$.}
\begin{equation}
\label{eq:S_mat_as_function_of_M}
S_{ab}(\theta_{p p'})=\frac{M_{ab}(\theta_{p p'})}{4 \hat{m}_a \hat{m}_b \sinh{\theta_{p p'}}}\,,
\end{equation}
where we assume $\theta_p$, $\theta_{p'} \in \mathbb{R}$ and $\theta_p>\theta_{p'}$.
The normalization $4 \hat{m}_a \hat{m}_b \sinh{\theta_{p p'}}$ comes from expressing the Dirac delta function of the overall energy-momentum conservation in terms of the rapidities
and dressing the external particles with their normalization factors. For the sake of clarity we report the derivation of formula~\eqref{eq:S_mat_as_function_of_M} in appendix~\ref{app:Jacobian}, even if it is a well-known result.
We also stress that the on-shell momenta of the external particles are parameterized in terms of the rapidities as follows
\begin{equation}
p = \hat{m}_a (\cosh{\theta_p}, \sinh{\theta_p}) \qquad , \qquad p' = \hat{m}_b (\cosh{\theta_{p'}}, \sinh{\theta_{p'}})\,,
\end{equation}
where $\hat{m}_a$ and $\hat{m}_b$ are the physical renormalized masses of the particles.

Universal expressions for these elastic amplitudes were obtained in~\cite{Fabri:2024qgd}, where it was found that
\begin{equation}
\begin{split}
&M^{(1\text{-loop})}_{ab}(\theta_{pp'})+ M^{(\text{ct.I})}_{a b}(\theta_{pp'})=\\
& \sum_{j \in \{\text{prop, ext} \}} \delta m_j^2 \frac{\partial}{\partial \hat{m}_j^2} M^{(0)}_{ab}(\theta_{p p'})-\frac{1}{2} \left( \frac{\delta m_a^2}{\hat{m}_a^2} +\frac{\delta m_b^2}{\hat{m}_b^2} \right) M^{(0)}_{ab}(\theta_{p p'} )\\
&+\frac{\bigl(M^{(0)}_{ab}(\theta_{p p'}) \bigl)^2}{8 \hat{m}_a \hat{m}_b \sinh{\theta_{p p'}}}+\frac{i \ai}{2 \pi} \hat{m}_a \hat{m}_b \sinh{\theta_{p p'}} \theta_{p p'} \frac{\partial}{\partial \theta_{p p'}} S^{(0)}_{ab}(\theta_{p p'}) \sum^r_{e=1} \hat{m}_e^2 \\
&+\frac{i}{\pi} \hat{m}_a \hat{m}_b \sinh{\theta_{pp'}}  \sum_{e=1}^r \frac{\partial}{\partial \theta_{p p'}} \ \text{p.v.} \int^{+\infty}_{-\infty} d\theta_k \,  S^{(0)}_{ae} (\theta_{pk} ) S^{(0)}_{b e} (\theta_{p' k}) \,.
\end{split}
\end{equation}
Again, using $\hat{m}_k$ or $m_k$ for the masses is irrelevant in the formula above since their difference starts at the two-loop order. 
The prescription is the same as the one used in~\cite{Fabri:2024qgd} and we summarize it below.
We define
\begin{equation}
\label{values_of_tree_level_amplitudes_at_infty}
\ai \equiv \frac{M^{(0)}_{i j} (\infty)}{m_i^2 m_j^2}
\end{equation}
to be the value of the tree-level amplitude at infinity rapidity normalized by the masses of the scattered particles. This number depends on the theory but not on the particles $i$ and $j$, as proven in~\cite{Fabri:2024qgd}. ``p.v.'' refers to the Cauchy principal value prescription, necessary to prevent collinear singularities at $\theta_k=\theta_p$ and $\theta_k=\theta_{p'}$ on the real integration line. This corresponds to performing the integral wrt to $\theta_k$ on the domain
\begin{equation}
(-\infty, \theta_{p'}-\epsilon) \cup (\theta_{p'}+ \epsilon,\theta_{p}- \epsilon) \cup (\theta_{p}+ \epsilon,+\infty)
\end{equation}
and sending $\epsilon \to 0$ after having integrated.
Adding to the expression above the coupling counterterms and the tree-level amplitude evaluated at the renormalized values of the masses we obtain
\begin{equation}
\begin{split}
&M^{(0)}_{a b}(\theta_{pp'})\Bigl|_{\hat{m}^2_j, \hat{C}^{(n)}}+M^{(1\text{-loop})}_{ab}(\theta_{pp'})+ M^{(\text{ct.I})}_{a b}(\theta_{pp'})+M^{(\text{ct.II})}_{a b}(\theta_{pp'})=\\
& M^{(0)}_{a b}(\theta_{pp'})\Bigl|_{\hat{m}^2_j + \delta m_j^2, \hat{C}^{(n)}+\delta C^{(n)}}-\frac{1}{2} \left( \frac{\delta m_a^2}{\hat{m}_a^2} +\frac{\delta m_b^2}{\hat{m}_b^2} \right) M^{(0)}_{ab}(\theta_{p p'} )\\
&+\frac{\bigl(M^{(0)}_{ab}(\theta_{p p'}) \bigl)^2}{8 \hat{m}_a \hat{m}_b \sinh{\theta_{p p'}}}+\frac{i \ai}{2 \pi} \hat{m}_a \hat{m}_b \sinh{\theta_{p p'}} \theta_{p p'} \frac{\partial}{\partial \theta_{p p'}} S^{(0)}_{ab}(\theta_{p p'}) \sum^r_{e=1} \hat{m}_e^2 \\
&+\frac{i}{\pi} \hat{m}_a \hat{m}_b \sinh{\theta_{pp'}}  \sum_{e=1}^r \frac{\partial}{\partial \theta_{p p'}} \ \text{p.v.} \int^{+\infty}_{-\infty} d\theta_k \,  S^{(0)}_{ae} (\theta_{pk} ) S^{(0)}_{b e} (\theta_{p' k}) \,.
\end{split}
\end{equation}
To derive the S-matrix from the expression above we need to multiply by the normalization factor in~\eqref{eq:S_mat_as_function_of_M}. Using that $\hat{m}^2_j + \delta m_j^2=m_j^2$ and $\hat{C}^{(n)}+\delta C^{(n)}=C^{(n)}$, as specified in~\eqref{eq:ren_masses_couplings}, the second row in the expression above (after normalising) becomes
\begin{equation}
\begin{split}
&\frac{1}{4 \hat{m}_a \hat{m}_b \sinh{\theta_{p p'}}} \biggl[ M^{(0)}_{a b}(\theta_{pp'})\Bigl|_{{m}^2_j , {C}^{(n)}}-\frac{1}{2} \left( \frac{\delta m_a^2}{\hat{m}_a^2} +\frac{\delta m_b^2}{\hat{m}_b^2} \right) M^{(0)}_{ab}(\theta_{p p'} ) \biggr]\\
&=\frac{1}{4 m_a m_b \sinh{\theta_{p p'}}}M^{(0)}_{a b}(\theta_{pp'})\Bigl|_{{m}^2_j , {C}^{(n)}} + \dots
\end{split}
\end{equation}
where we used that
\begin{equation}
\hat{m}_j=\sqrt{m_j^2-\delta m^2_j}=m_j \left(1-\frac{\delta m^2_j}{2 m_j^2}+ \dots \right)\,.
\end{equation}
As before the ellipses contain terms contributing at two loops and higher orders.
Then the renormalized S-matrix to one loop order in perturbation theory is given by
\begin{equation}
S_{ab}(\theta_{pp'})=S^{(0)}_{ab}(\theta_{pp'})+S^{(1)}_{ab}(\theta_{pp'})
\end{equation}
where we defined
\begin{equation}
S^{(0)}_{ab}(\theta_{pp'})=\frac{1}{4 m_a m_b \sinh{\theta_{pp'}}} \ M^{(0)}_{a b}(\theta_{pp'})\Bigl|_{{m}^2_j , {C}^{(n)}}
\end{equation}
and
\begin{equation}
\label{eq:result_S_mat_eq_m_ren_2Int}
\begin{split}
S^{(1)}_{ab}(\theta_{pp'})&=\frac{\bigl(S^{(0)}_{ab}(\theta_{p p'}) \bigl)^2}{2}+\frac{i \ai}{8 \pi}  \theta_{p p'} \frac{\partial}{\partial \theta_{p p'}} S^{(0)}_{ab}(\theta_{p p'}) \sum^r_{e=1} m_e^2 \\
&+\frac{i}{4 \pi}  \sum_{e=1}^r \frac{\partial}{\partial \theta_{p p'}} \ \text{p.v.} \int^{+\infty}_{-\infty} d\theta_k \   S^{(0)}_{ea} (\theta_{k p} ) S^{(0)}_{e b} (\theta_{k p'}) \,.
\end{split}
\end{equation}
The expression~\eqref{eq:result_S_mat_eq_m_ren_2Int} is identical to equation (2.23) in~\cite{Fabri:2024qgd} and we just showed that it is valid not only on models having mass ratios which do not renormalise under one-loop corrections (as specified in~\cite{Fabri:2024qgd}) but to any model with a classical integrable Lagrangian of type~\eqref{eq0_1} with physical masses and couplings defined by \eqref{eq:ren_masses_couplings}. 

We also stress that the expression for one-loop S-matrices reported above comes from summing over all one-loop Feynman diagrams and counterterms. The way in which we wrote the final answer for the S-matrix only comes from rewriting this sum of Feynman diagrams in a way in which combinations of diagrams are expressed in terms of on-shell tree-level amplitudes (see formula~\eqref{eq:splitting_ono_loop_amplitude_into_double_and_single_cuts} and associated discussion). While in inelastic processes most of the tree-level amplitudes appearing in~\eqref{eq:splitting_ono_loop_amplitude_into_double_and_single_cuts} cancel, for elastic processes they generate the expression in~\eqref{eq:result_S_mat_eq_m_ren_2Int}~\cite{Fabri:2024qgd}.
As a consequence of this fact, the expression in~\eqref{eq:result_S_mat_eq_m_ren_2Int} is the complete one-loop S-matrix, carrying information both on simple poles and anomalous thresholds. 
An explanation on the generation of anomalous thresholds in elastic processes starting from~\eqref{eq:result_S_mat_eq_m_ren_2Int} can be found in~\cite{Fabri:2024qgd}.
In the same paper, it was shown how to write the last line in~\eqref{eq:result_S_mat_eq_m_ren_2Int} as a closed integral in the complex rapidity plane using the property 
\begin{equation}
\label{eq:periodicity_tree_level_S}
S^{(0)}_{ea}(\theta)=-S^{(0)}_{\bar{e}a}(\theta+i \pi)\,,
\end{equation}
which can be explicitly checked at the level of tree-level Feynman diagrams\footnote{This comes from noting that under the transformation $\theta \to \theta+i\pi$ the $s$ and $t$ Mandelstam variables, which in elastic processes are defined by
\begin{equation}
s=m_a^2+m_e^2+2 m_a m_e \cosh{\theta}\,, \qquad t=m_a^2+m_e^2-2 m_a m_e \cosh{\theta}\,,
\end{equation}
are exchanged. This is similar to a crossing transformation, which instead is implemented by sending $\theta \to i \pi - \theta$. The minus sign in~\eqref{eq:periodicity_tree_level_S} arises from the normalization factor in~\eqref{eq:S_mat_as_function_of_M}.}.
Then following~\cite{Fabri:2024qgd} the expression in~\eqref{eq:result_S_mat_eq_m_ren_2Int}  can be written as
\begin{equation}
\label{eq:result_S_mat_eq_m_ren_3}
\begin{split}
S^{(1)}_{ab}(\theta_{pp'})&=\frac{\bigl(S^{(0)}_{ab}(\theta_{p p'}) \bigl)^2}{2}+\frac{i \ai}{8 \pi}  \theta_{p p'} \frac{\partial}{\partial \theta_{p p'}} S^{(0)}_{ab}(\theta_{p p'}) \sum^r_{e=1} m_e^2 \\
&-\frac{1}{4n \pi^2}  \sum_{e=1}^r \frac{\partial}{\partial \theta_{p p'}} \oint_{\ip_n} d\theta_k \theta_k S^{(0)}_{ea} (\theta_{k p} ) S^{(0)}_{e b} (\theta_{k p'})\\
&+\frac{1}{16}  \left( \frac{M^{(0)}_{aa}(0)}{m_a^2}- \frac{M^{(0)}_{bb}(0)}{m_b^2} \right) \frac{\partial}{\partial \theta_{p p'}} S^{(0)}_{ab}(\theta_{p p'}) \,,
\end{split}
\end{equation}
where $\ip_n$ is a rectangle in the complex $\theta_k$ plane having infinitely long horizontal sides on the lower edge of the real line and on the lower edge of the line with a constant imaginary part equal to $\Im(n \pi)$. The rectangular path follows the counterclockwise direction.
Once again it is remarkable that the mass shifts $\{\delta m^2_j\}^r_{j=1}$ \emph{completely disappear} from the final formula \eqref{eq:result_S_mat_eq_m_ren_3} for the one-loop S-matrices although they are essential to define the physical masses.

Let us also stress that in 1+1 dimensional theories with no production crossing symmetry takes the following simple form
$$
S_{ab}(\theta)=S_{\bar{a}b}(i \pi-\theta)
$$
in the rapidity plane and must hold order by order in the perturbative expansion
$$
S_{ab}(\theta)=1+S^{(0)}_{ab}(\theta)+S^{(1)}_{ab}(\theta)+ \dots \,.
$$
It would be interesting to check if starting with a tree-level crossing symmetric set of S-matrix elements $S^{(0)}_{ab}(\theta)$ our formula implies that $S^{(1)}_{ab}(\theta)$ must also satisfy crossing. We checked that this is the case for all examples of models considered; however, we did not manage to prove this fact universally for general S-matrices.

\subsection{Mass corrections and double poles in inelastic amplitudes}
\label{sec:mass-corrections-inelastic-amp}

It is a well-known fact that one-loop amplitudes of 1+1 dimensional quantum field theories can have, in addition to simple poles due to the propagation of bound states, also second-order poles. These second-order poles can be explained from a perturbative point of view in terms of anomalous thresholds due to multiple propagators going simultaneously on-shell in collections of one-loop Feynman diagrams~\cite{Coleman:1978kk,Braden:1990wx}. These thresholds are special cases of Landau singularities.

Differently from four dimensions, where these thresholds normally appear as branch cuts (we remand to~\cite{Fevola:2023fzn} for a historical overview of the topic in higher dimensional theories), in two dimensions the dimensionality of the space makes these thresholds double poles. As highlighted in~\cite{Coleman:1978kk} this is not related with the particular two dimensional theory under consideration (whether it is integrable or not); these double poles can appear in any two-dimensional theory with a suitable choice of the mass ratios.

Of course, for an inelastic process to vanish the residue of the total amplitude at these double poles must be zero. 
In this subsection, we show that these thresholds in one-loop inelastic amplitudes can be computed through formula~\eqref{ab_to_cd_one_loop_inelastic_amp_derivatives}.

We find that the coefficients of these second-order poles in one-loop inelastic amplitudes vanish only if all masses scale equally under one-loop quantum corrections, which is
\begin{equation}
\label{eq:equal_scaling_masses}
\delta m_j^2= \gamma \, m_j^2\,, \qquad j=1, \dots, r\,,
\end{equation}
with $\gamma$ being independent of the particle label $j$. In models not satisfying condition~\eqref{eq:equal_scaling_masses} these thresholds do not vanish by themselves in one-loop inelastic amplitudes; instead they are cancelled
against the expansion of the tree-level amplitude around the values of the classical masses, namely
\begin{equation}
\label{eq:cancel_double_poles_expanding_tree_amplitude}
\begin{split}
&M^{(0)}_{ab \to cd}\Bigl|_{\hat{m}^2_k, C^{(n)}}+ \sum_{k \in \{\text{prop}, \text{ext}\}} \delta m_k^2 \frac{\partial}{\partial \mu^2_k} M_{ab \to cd}^{(0)}\Bigl|_{\mu^2_k=\hat{m}^2_k}\\
&=M^{(0)}_{ab \to cd}\Bigl|_{m^2_k, C^{(n)}}\\
&-  \sum_{k \in \{\text{prop}, \text{ext}\}} \delta m_k^2 \frac{\partial}{\partial \mu^2_k} M_{ab \to cd}^{(0)}\Bigl|_{\mu^2_k=\hat{m}^2_k}+ \sum_{k \in \{\text{prop}, \text{ext}\}} \delta m_k^2 \frac{\partial}{\partial \mu^2_k} M_{ab \to cd}^{(0)}\Bigl|_{\mu^2_k=\hat{m}^2_k}=0\,.
\end{split}
\end{equation}
The total amplitude (composed of the tree-level and one-loop amplitude) is then always zero by construction, even though Landau diagrams do not sum to zero in one-loop inelastic processes.

We start computing explicitly the expression in \eqref{ab_to_cd_one_loop_inelastic_amp_derivatives}, corresponding to the sum over all connected one-loop diagrams and tree-level diagrams containing two-point counterterms. As previously mentioned, we consider $\theta_{pp'}$ to be free and $\{\theta_{pq'}, \theta_{pq} \}$ to be fixed by momentum conservation and thus dependent on $\theta_{pp'}$ and external masses. Due to this, writing $s$ as in~\eqref{eq:s_t_u_Mandelstam}, we obtain 
\begin{multline}
        \sum_{k \in \{\text{prop}, \text{ext}\}} \delta m_k^2 \frac{\partial}{\partial \mu^2_k} \frac{1}{s-\mu_i^2} \Bigl|_{\mu^2_k=m^2_k} \\
        = - \left( \frac{\delta m_a^2}{2m_a} \frac{\partial s} {\partial \mu_a}\Bigl|_{\mu_a=m_a} + \frac{\delta m_b^2}{2m_b} \frac{\partial s} {\partial \mu_b}\Bigl|_{\mu_b=m_b} - \delta m_i^2 \right) \ \frac{1}{(s-m_i^2)^2}.
\end{multline}
In this case $\theta_{pp'}$ is a free variable and does not contain any dependence on the masses; then we can directly compute the derivatives of $s$ with respect to the masses by using \eqref{eq:s_t_u_Mandelstam} and we find the following expression
\begin{equation}
\sum_{j \in \{\text{prop}, \text{ext}\}} \delta m_j^2 \frac{\partial}{\partial \mu^2_j} \frac{1}{s-\mu_i^2} \Bigl|_{\mu^2_j=m^2_j} = \frac{\mathcal{F}^{(i)}_s(s)}{(s-m_i^2)^2} \,,
\end{equation}
where 
\begin{equation}
\label{Fischannel}
\mathcal{F}^{(i)}_s(s) \equiv \delta m_i^2 - \frac{\delta m_a^2}{m_a^2} \frac{(s+m_a^2 - m_b^2)}{2} - \frac{\delta m_b^2}{m_b^2} \frac{(s + m_b^2-m_a^2)}{2}.
\end{equation}
Differently from the Mandelstam variable $s$, working out the derivatives of $t$ and $u$ with respect to the external masses is not completely straightforward since $\theta_{pq'}$ and $\theta_{pq}$ depend on $\theta_{pp'}$ and the masses of the external particles.
To obtain these derivatives we assume the rapidities to be purely imaginary numbers (this is indeed the region where the poles are located) and use the results of appendix \ref{app:geometrical_relations}. In particular, we can read the derivatives of $t$ and $u$ from \eqref{eq:dsdtrelation_appendix} and \eqref{eq:dsdurelation_appendix}, after requiring that $s$ is also a function of the external masses through~\eqref{eq:s_t_u_Mandelstam}. From~\eqref{eq:dsdtrelation_appendix} we have  
\begin{align}
\label{eq:t-der1}
    & \frac{\partial t}{\partial \mu_a^2} = \frac{\Delta_{tbc}}{\Delta_{sab}} - \frac{\Delta_{tbc}\Delta_{tad}}{\Delta_{sab}\Delta_{scd}} \frac{\partial s}{\partial \mu_a^2}, \ \ \ \ \  \frac{\partial t}{\partial \mu_b^2} = \frac{\Delta_{tad}}{\Delta_{sab}} - \frac{\Delta_{tbc}\Delta_{tad}}{\Delta_{sab}\Delta_{scd}} \frac{\partial s}{\partial \mu_b^2}, \\
    \label{eq:t-der2}
    & \frac{\partial t}{\partial \mu_c^2} = \frac{\Delta_{tad}}{\Delta_{scd}}, \hspace{3.45cm} \frac{\partial t}{\partial \mu_d^2} = \frac{\Delta_{tbc}}{\Delta_{scd}},
\end{align}
where $\Delta_{sij}$ ($\Delta_{tij}$) denotes the area of the triangles with sides $m_i$, $m_j$ and $\sqrt{s}$ ($\sqrt{t}$). 
Similarly, for the $u$ derivatives, it holds that
\begin{align}
\label{eq:u-der1}
    & \frac{\partial u}{\partial \mu_a^2} = \frac{\Delta_{ubd}}{\Delta_{sab}} + \frac{\Delta_{uac}\Delta_{ubd}}{\Delta_{scd}\Delta_{sab}} \frac{\partial s}{\partial \mu_a^2}, \ \ \ \ \  \frac{\partial u}{\partial \mu_b^2} =  \frac{\Delta_{uac}}{\Delta_{sab}} + \frac{\Delta_{uac}\Delta_{ubd}}{\Delta_{scd}\Delta_{sab}} \frac{\partial s}{\partial \mu_b^2}, \\
    \label{eq:u-der2}
    & \frac{\partial u}{\partial \mu_c^2} = -\frac{\Delta_{ubd}}{\Delta_{scd}}, \hspace{3.22cm} \frac{\partial u}{\partial \mu_d^2} = -\frac{\Delta_{uac}}{\Delta_{scd}} \,.
\end{align}
By making use of the derivatives above we find that 
\begin{align}
&\sum_{k \in \{\text{prop}, \text{ext}\}} \delta m_k^2 \frac{\partial}{\partial \mu^2_k} \frac{1}{t-\mu_j^2} \Bigl|_{\mu^2_k=m^2_k} = \frac{\mathcal{F}^{(j)}_t (s,t)}{(t-m_j^2)^2}\,,\\
    & \sum_{k \in \{\text{prop}, \text{ext}\}} \delta m_k^2 \frac{\partial}{\partial \mu^2_k} \frac{1}{u-\mu_l^2} \Bigl|_{\mu^2_j=m^2_j} = \frac{\mathcal{F}_u^{(l)} (s,u)}{(u-m_l^2)^2} \,,
\end{align}
where for the $t$ channel we have
\begin{equation}
\label{Fjtchannel}
\begin{split}
    & \mathcal{F}^{(j)}_t (s,t) \equiv \delta m_j^2 - \delta m_a^2 \left( \frac{\Delta_{tbc}}{\Delta_{sab}} - \frac{\Delta_{tbc}\Delta_{tad}}{\Delta_{sab} {\Delta_{scd}}} \frac{(s  + m_a^2-m_b^2)}{2 m_a^2} \right) \\
    &  - \delta m_b^2 \left( \frac{\Delta_{tad}}{\Delta_{sab}} - \frac{\Delta_{tbc}\Delta_{tad}}{\Delta_{sab} {\Delta_{scd}}} \frac{(s + m_b^2 - m_a^2)}{2 m_b^2} \right)- \delta m_c^2 \frac{\Delta_{tad}}{\Delta_{scd}} - \delta m_d^2 \frac{\Delta_{tbc}}{\Delta_{scd}}
    \end{split}
\end{equation}
and for the $u$ channel we end up with
\begin{equation}
\label{Fluchannel}
\begin{split}
    &\mathcal{F}^{(l)}_u(s,u) \equiv \delta m_l^2 - \delta m_a^2 \left(  \frac{\Delta_{ubd}}{\Delta_{sab}} + \frac{\Delta_{uac}\Delta_{ubd}}{\Delta_{sab}\Delta_{scd}} \frac{(s + m_a^2- m_b^2)}{2 m_a^2} \right) \\
    & - \delta m_b^2 \left( \frac{\Delta_{uac}}{\Delta_{sab}} + \frac{\Delta_{uac}\Delta_{ubd}}{\Delta_{sab}\Delta_{scd}} \frac{(s + m_b^2 - m_a^2)}{2m_b^2} \right)
     + \delta m_c^2 \frac{\Delta_{ubd}}{\Delta_{scd}} + \delta m_d^2 \frac{\Delta_{uac}}{\Delta_{scd}}\,.
\end{split}
\end{equation}
In the end, we can write \eqref{ab_to_cd_one_loop_inelastic_amp_derivatives} as 
\begin{multline}
\label{Mab_to_cd_plus_ctI_dpoles}
    M_{ab\rightarrow cd}^{\textrm{(1-loop)}} + M_{ab\rightarrow cd}^{\textrm{(ct.I)}} = -i \sum_{i\in s} C^{(3)}_{ab\bar{i}} C^{(3)}_{i\bar{c}\bar{d}} \frac{\mathcal{F}^{(i)}_s (s)}{(s-m_i^2)^2} 
    -i \sum_{j\in t} C^{(3)}_{a\bar{d}\bar{j}} C^{(3)}_{jb\bar{c}} \frac{\mathcal{F}^{(j)}_t (s,t)}{(t-m_j^2)^2} \\
    -i \sum_{l\in u} C^{(3)}_{a\bar{c}\bar{l}} C^{(3)}_{lb\bar{d}} \frac{\mathcal{F}^{(l)}_u (s,u)}{(u-m_l^2)^2}.
\end{multline}
We notice that, in general, the expression above has double poles corresponding to the situations where $s=m_i^2$, $t=m_j^2$ or $u=m_l^2$, for some particle $i$, $j$ and $l$ propagating in the $s$-, $t$- or $u$-channel.
These double-poles must be identified with Landau singularities generated by one-loop diagrams contributing to inelastic processes. 

In section~\ref{sec_Landau_poles_inel_processes} we will compare explicitly the coefficient at the second-order poles obtained from the expression above with Landau diagrams. For now, let us see under which conditions on the mass radiative corrections these double poles cancel. 
For a tree-level inelastic amplitude $M_{ab\rightarrow cd}^{\textrm{(0)}}$ to vanish it must hold that at the pole position $s=m_i^2$, there are simultaneous singularities also in the $t$ and $u$ channels, for some internal on-shell propagating particles of types $j$ and $l$; these particles are also on-shell, that is
\begin{equation}
t=m_j^2\qquad \textrm{and} \qquad u=m_l^2\,.
\end{equation}
We remand the reader to appendix~\ref{app:tree_canc_and_flipping} for a short review of how these singularities cancel in tree-level processes, while a more detailed discussion can be found in~\cite{Dorey:2021hub}.
While for suitable $3$-point couplings
the condition just mentioned is enough for the cancellation of the tree-level amplitude $M_{ab\rightarrow cd}^{\textrm{(0)}}$, it is not enough to ensure the cancellation of~\eqref{Mab_to_cd_plus_ctI_dpoles}, as we will see in one moment.
To compute the double pole residues we expand $t$ and $u$ at the pole as function of $s$ using the following formulas~\cite{Dorey:2021hub}
\begin{equation}
    \frac{d t}{d s}\Bigl|_{s=m_i^2} = - \frac{\Delta_{jcb}\Delta_{jda}}{\Delta_{iba}\Delta_{idc}} \ \ \ \ \textrm{and} \ \ \ \ \frac{d u}{d s}\Bigl|_{s=m_i^2} =  \frac{\Delta_{lac}\Delta_{lbd}}{\Delta_{icd}\Delta_{iab}}.
\end{equation}
These formulas are obtained from \eqref{eq:dsdtrelation_appendix} and \eqref{eq:dsdurelation_appendix} for variations with fixed external masses, i.e., $d \mu_j^2 =0 $. 
Assuming no degeneracy and parameterising the couplings as in~\eqref{eq:417068}
the double pole residue at $s=m_i^2$ is given by
\begin{equation}
\label{eq:residue_pole_oneloop_generic_1st}
\begin{split}
    &M_{ab\rightarrow cd}^{\textrm{(1-loop)}} + M_{ab\rightarrow cd}^{\textrm{(ct.I)}} \sim - \frac{i}{(s-m_i^2)^2}  \Delta^2_{ab i} \Delta^2_{icd}\\
    &\times \biggl[\frac{f_{\bar{i}ab} f_{i\bar{c}\bar{d}}}{\Delta_{iab} \Delta_{icd}}\ \mathcal{F}^{(i)}_s (m_i^2)
    + \frac{f_{jb\bar{c}} f_{\bar{j} a\bar{d}} }{\Delta_{jbc}\Delta_{jad}}\ \mathcal{F}^{(j)}_t (m_i^2,m_j^2)
    +   \frac{f_{\bar{l} a\bar{c}} f_{lb\bar{d}}}{\Delta_{lac}\Delta_{lbd}} \ \mathcal{F}^{(l)}_u (m_i^2,m_l^2) \biggr] \,.
    \end{split}
\end{equation}
The vanishing of the residue leads to the following constraint
\begin{equation}
\label{eq:residue_pole_oneloop_generic}
    \frac{f_{\bar{i} ab} f_{i\bar{c}\bar{d}}}{\Delta_{iab} \Delta_{icd}}\ \mathcal{F}^{(i)}_s (m_i^2)
    + \frac{f_{jb\bar{c}} f_{\bar{j} a\bar{d}}}{\Delta_{jbc}\Delta_{jad}}\ \mathcal{F}^{(j)}_t (m_i^2,m_j^2)
    +   \frac{f_{\bar{l} a\bar{c}} f_{lb\bar{d}}}{\Delta_{lac}\Delta_{lbd}} \ \mathcal{F}^{(l)}_u (m_i^2,m_l^2) = 0\,.
\end{equation}
Equation~\eqref{eq:residue_pole_oneloop_generic} is the one-loop generalization of the tree-level result~\eqref{eq:94754}; as in~\eqref{eq:94754}, the constraint above can be readily generalized to the presence of degeneracies or when only two diagrams are simultaneously singular. In the latter case, we remove the term corresponding to the non-singular diagram from~\eqref{eq:residue_pole_oneloop_generic}.
If condition~\eqref{eq:residue_pole_oneloop_generic} is satisfied then the amplitude~\eqref{ab_to_cd_one_loop_inelastic_amp_derivatives} can contain at most simple poles.

\begin{figure}
\begin{center}
\includegraphics*[width=0.8\textwidth]{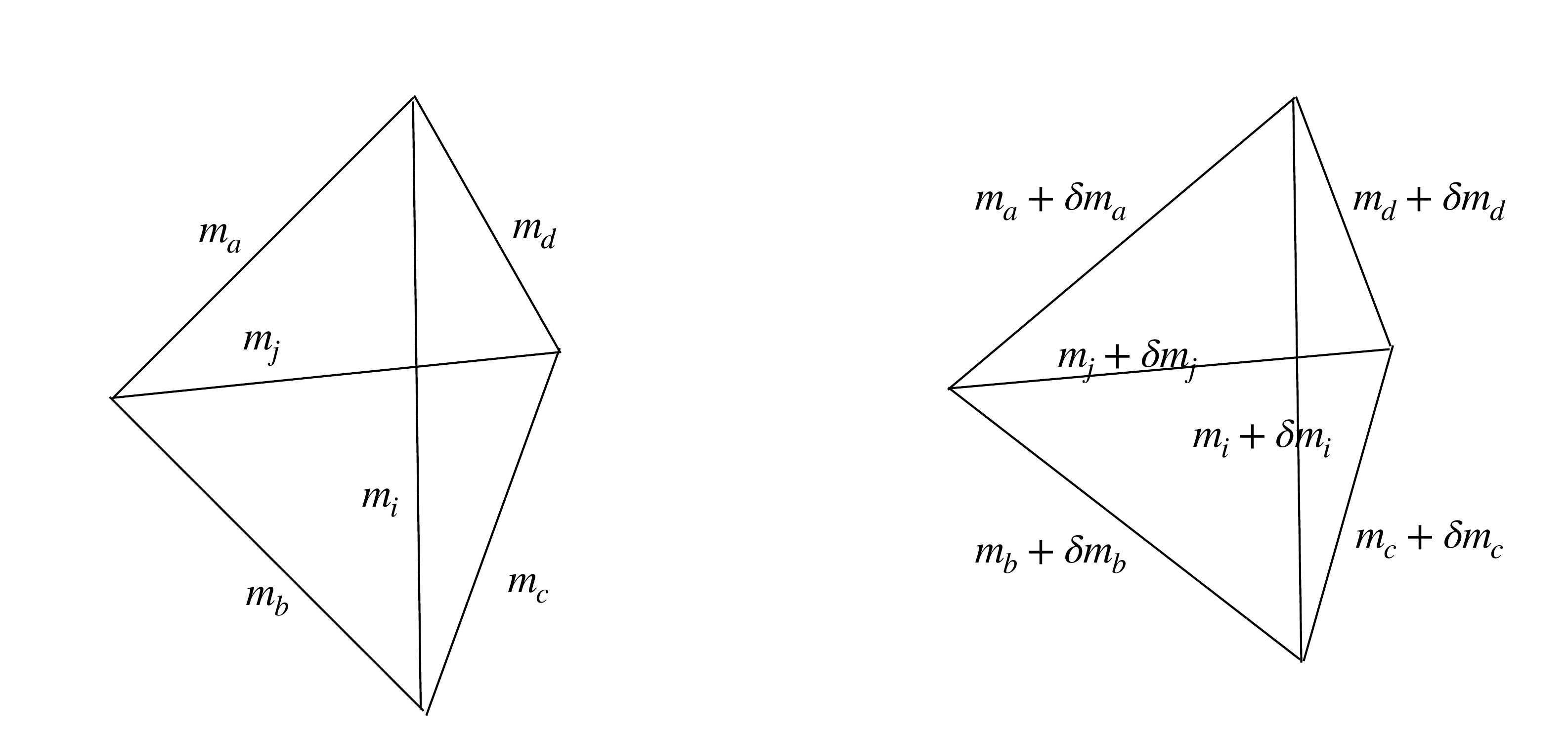} 
\end{center}
\caption{On the left, we have a pair of overlapped on-shell diagrams contributing to the tree-level amplitude~\eqref{eq:160215} and on the right the same pair of diagrams with deformed masses. Both quadrilaterals above are planar.}
\label{quadrilateral_variation_flipping}
\end{figure}

Let us assume that the on-shell bound states appear only in the $s$ and $t$ channels, and we have no singularity in the $u$ channel.
Then using~\eqref{eq:94754}, with $f_{a\bar{c}\bar{l}} f_{lb\bar{d}}=0$, the constraint~\eqref{eq:residue_pole_oneloop_generic} becomes
\begin{equation}
 \frac{\mathcal{F}^{(i)}_s (m_i^2)}{\Delta_{iab}\Delta_{icd}}\ 
    + \frac{\mathcal{F}^{(j)}_t (m_i^2,m_j^2)}{\Delta_{jbc}\Delta_{jad}}=0\, .
\end{equation}
Using~\eqref{Fischannel} and~\eqref{Fjtchannel}, it is simple to check that the expression above implies that
\begin{equation}
\label{eq:st_cancellation_deltam}
\delta m_j^2= \delta m^2_a \frac{\Delta_{jbc}}{\Delta_{iab}}+\delta m^2_b \frac{\Delta_{jad}}{\Delta_{iab}}+\delta m^2_c \frac{\Delta_{jad}}{\Delta_{icd}}+\delta m^2_d \frac{\Delta_{jbc}}{\Delta_{icd}} - \delta m^2_i \frac{\Delta_{jad} \Delta_{jbc}}{\Delta_{icd} \Delta_{iab}} \quad \text{($s$-$t$ poles)}\,.
\end{equation}
This condition for the cancellation of the double pole is equal to~\eqref{eq:dsdtrelation_appendix} after substituting $s$ and $t$ with the squares of the on-shell masses $m^2_i$ and $m^2_j$ and (as~\eqref{eq:dsdtrelation_appendix}) is a constraint relating distances between points in $\mathbb{R}^2$. In particular the meaning of equation~\eqref{eq:st_cancellation_deltam} is the following: given 
\begin{equation}
\label{eq:massesabcdij}
\{m_a, \, m_b, \, m_c, \, m_d, \, m_i, \,m_j \}
\end{equation}
to be distances between points in $\mathbb{R}^2$ (as depicted on the LHS of figure~\ref{quadrilateral_variation_flipping}) then 
\begin{equation}
\label{eq:defmassesabcdij}
\{m_a+ \delta m_a, \,  m_b+\delta m_b, \, m_c+\delta m_c, \, m_d+\delta m_d, \, m_i+\delta m_i, \,m_j+\delta m_j \}
\end{equation}
must also correspond to distances between points in $\mathbb{R}^2$ (see the RHS of figure~\ref{quadrilateral_variation_flipping}).
This is a necessary condition for the cancellation of the double pole when simultaneous singularities appear in the $s$ and $t$ channels and is equivalent to the requirement for the radiative corrections to the masses to not affect the flipping rule in the $s/t$ channels. In other words 
if the masses in~\eqref{eq:massesabcdij} form a tiled planar quadrilateral of the form on the LHS in figure~\ref{quadrilateral_variation_flipping}, then the deformed masses in~\eqref{eq:defmassesabcdij} must also form a tiled planar quadrilateral, as depicted on the RHS of figure~\ref{quadrilateral_variation_flipping}.
This means that if exists a pair of simultaneously singular cancelling diagrams in the tree-level amplitude $M^{(0)}_{ab \to cd}$ then the two diagrams must be simultaneously singular also after having received the mass corrections $\delta m_k$ (with $k= a,b,c,d,i,j$).

Analogously if we consider the situation in which the singularities happen in the $s$-$u$ or in the $t$-$u$ channels, we find the constraints
\begin{equation}
\label{eq:su_cancellation_deltam}
\delta m^2_l= \delta m^2_a \frac{\Delta_{lbd}}{\Delta_{iab}}+\delta m^2_b \frac{\Delta_{lac}}{\Delta_{iab}}-\delta m^2_c \frac{\Delta_{lbd}}{\Delta_{icd}} -\delta m^2_d \frac{\Delta_{lac}}{\Delta_{icd}} + \delta m_i^2 \frac{\Delta_{lac} \Delta_{lbd}}{\Delta_{icd} \Delta_{iab}} \quad \text{($s$-$u$ poles)}
\end{equation}
and
\begin{equation}
\label{eq:tu_cancellation_deltam}
\frac{\delta m_j^2}{\Delta_{jad} \Delta_{jbc}}+\frac{\delta m_l^2}{\Delta_{lac} \Delta_{lbd}}=\frac{\delta m_a^2}{\Delta_{lac} \Delta_{jad}}+\frac{\delta m_b^2}{\Delta_{jbc} \Delta_{lbd}}-\frac{\delta m_c^2}{\Delta_{lac} \Delta_{jbc}}-\frac{\delta m_d^2}{\Delta_{lbd} \Delta_{jad}}
\quad \text{($t$-$u$ poles)}\,,
\end{equation}
with $i$, $j$ and $l$ being the particles propagating in the $s$, $t$ and $u$ channels, respectively.
The two constraints above correspond to the preservation of the flipping rule for pairs of tree-level cancelling diagrams in the $s/u$ and $t/u$ channels.
For a generic classically integrable theory of type~\eqref{eq0_1}, the only possible mass corrections preserving the flipping rule for all processes should correspond to the trivial solution~\eqref{eq:equal_scaling_masses}. Note that~\eqref{eq:equal_scaling_masses} is a dilatation and preserves the flipping rule trivially since it preserves all the fusing angles of the starting classically integrable theory. 

If there were non-trivial solutions to the set of constraints~\eqref{eq:st_cancellation_deltam}, \eqref{eq:su_cancellation_deltam} and~\eqref{eq:tu_cancellation_deltam} then for a given classically integrable theory of type~\eqref{eq0_1} it would be possible to change its masses continuously and find another classically integrable theory. For doing this it would be enough to choose the deformation of the masses in such a way as to preserve the flipping rule, along the direction indicated by $\{\delta m_k\}^r_{k=1}$.
However, it is a known fact that the mass ratios of classically integrable theories are discrete sets of numbers, which cannot be modified continuously.
For this reason, we expect no solutions to the constraints~\eqref{eq:st_cancellation_deltam}, \eqref{eq:su_cancellation_deltam} and~\eqref{eq:tu_cancellation_deltam} other than the dilatation of the classical masses. 

From the argument above, we expect that the only theories having no double poles in one-loop inelastic amplitudes are those in which condition~\eqref{eq:equal_scaling_masses} is satisfied (famous examples of these models are the self-dual affine Toda theories~\cite{Braden:1989bu}). For all the remaining theories we expect double poles in one-loop inelastic amplitudes; these poles are cancelled in the total amplitude only by choosing the renormalized masses to be different from the classical ones and expanding the tree-level amplitudes around the classical values of the masses (see \eqref{eq:cancel_double_poles_expanding_tree_amplitude}).
Due to this fact, the renormalized masses are coupling-dependent.

On the other hand, if condition~\eqref{eq:equal_scaling_masses} is satisfied then
we can in principle identify the renormalized masses with the classical ones and have these to be independent of the coupling. In this case, double poles cancel from one-loop amplitudes and we can introduce properly tuned $3$-point couplings to also cancel the simple poles and make all inelastic amplitudes vanish, as worked out in~\cite{Polvara:2023vnx, Fabri:2024qgd} for theories having mass ratios that do not scale under one-loop corrections.
However, this choice for the physical values of the masses is possible only if condition~\eqref{eq:equal_scaling_masses} is satisfied; in all the other cases we must require that the physical masses differ from the classical ones through~\eqref{eq:ren_masses_couplings} which makes them coupling dependent. This second approach is more general and is valid even if the masses do not scale equally; for this reason, we must assume this condition for the renormalized masses.

\section{One-loop S-matrices of nonsimply-laced models}
\label{OneloopS_mat_nonsimplaced}

Affine Toda models provide a well-known class of 1+1 dimensional massive quantum field theories which are classically integrable~\cite{Mikhailov:1980my,Olive:1984mb}. Their integrability is due to a Lax formulation of the equations of motion~\cite{Olive:1984mb} through which an infinite tower of higher-spin local conserved charges can be constructed. 
Assuming these charges survive the quantization, exact S-matrices have been advanced both for simply-~\cite{Arinshtein:1979pb,Freund:1989jq,Destri:1989pg,Christe:1989ah,Christe:1989my,Klassen:1989ui,Braden:1989bg,Braden:1989bu,Dorey:1990xa,Dorey:1991zp,Fring:1991gh} and nonsimply-laced theories~\cite{Delius:1991cu,Delius:1991kt,Corrigan:1993xh,Dorey:1993np,Oota:1997un}\footnote{Affine Toda theories containing fermionic fields have also been studied in the past~\cite{Delius:1990ij,Delius:1991sv} but will not be considered in this paper.}. Simply-laced affine Toda theories are the simplest examples of quantum integrable theories of type~\eqref{eq0_1} since they have mass ratios which are not affected by one-loop radiative corrections~\cite{Braden:1989bu,Christe:1989my}.
Despite the S-matrices of these theories being confirmed in the past through different perturbative checks~\cite{Braden:1990wx,Braden:1991vz,Braden:1990qa,Sasaki:1992sk,Braden:1992gh,Dorey:2022fvs,Dorey:2023cuq} based on a model-by-model study, only recently it was shown that these S-matrices match universally (both at tree level~\cite{Dorey:2021hub} and one loop~\cite{Fabri:2024qgd}) with the formulas originally proposed in~\cite{Dorey:1990xa}.

We now move to check the perturbative integrability of nonsimply-laced models which possess complicated dynamics by the fact that each mass gets a proper correction, as discussed in~\cite{Delius:1991cu,Delius:1991kt,Corrigan:1993xh}. 
This suggests either that the integrability of nonsimply-laced theories is broken at the quantum level, or that (if preserved) its link with the classical integrability is much less clear. This second option was proposed originally in~\cite{Delius:1991cu,Delius:1991kt}, where S-matrices of nonsimply-laced theories were proposed with physical-strip poles moving as functions of the coupling constant, according to the radiative corrections of the masses. In particular, it was advanced the idea that each dual-pair of Dynkin diagrams (i.e. diagrams mapping one into the other under the root transformation $\alpha_i \to 2 \alpha_i /\alpha_i^2$)
corresponds to a unique quantum theory having an S-matrix interpolating at weak and strong couplings between the 
tree-level S-matrices constructed from the affine Toda Lagrangians of the two dual Dynkin diagrams. We list below the four different classes of dual pairs of Dynkin diagrams
\begin{equation}
\begin{split}
&g^{(1)}_{2} \leftrightarrow d^{(3)}_4\,, \qquad f^{(1)}_4 \leftrightarrow e^{(2)}_6\,, \qquad c^{(1)}_{n} \leftrightarrow d^{(2)}_{n+1}\,, \qquad b^{(1)}_{n} \leftrightarrow a^{(2)}_{2n-1}\,,
\end{split}
\end{equation}
each associated with the quantum S-matrix of a nonsimply-laced affine Toda theory. The only exceptional class is $a^{(2)}_{2n}$, which is composed of self-dual models and has mass ratios which are not affected by quantum corrections.

In the following, we study one-loop processes in nonsimply-laced models and determine their one-loop S-matrices in terms of tree-level quantities using formula~\eqref{eq:result_S_mat_eq_m_ren_3}. 
As shown in the previous section, these models are one-loop integrable provided the renormalized masses are functions of the coupling as given by \eqref{eq:ren_masses_couplings} and we obtain that their one-loop S-matrices obtained from~\eqref{eq:result_S_mat_eq_m_ren_3} match exactly the results of~\cite{Delius:1991kt,Corrigan:1993xh}.

\subsection{Affine Toda field theories}
\label{sec:affine-toda}

Affine Toda field theories describe the interactions of $r$ massive scalar bosons $(\phi_1, \dots, \phi_r)$ in 1+1 dimensions through the following Lagrangian
\begin{equation}
\label{eq:lagrangian-toda}
\mathcal{L}=\frac{\partial_\mu \phi_a \partial^\mu \phi_a}{2} - \frac{\mu^2}{\g^2} \sum^r_{i=0} n_i e^{\g \, \alpha^a_i \phi_a}\,,
\end{equation}
where $\mu$ is a mass scale and $g$ is the coupling constant. The set of vectors $\{\alpha_0,\, \alpha_1,\, \dots,\, \alpha_r\} \in \mathbb{R}^r$ makes up the extended root system of an affine Lie algebra $\mathfrak{g}$ while the integers $n_i$ are the Kac labels and satisfy the condition
\begin{equation}
\sum_{i=0}^r n_i \alpha_i=0\,,
\end{equation}
with $n_0=1$.  This condition is necessary for having a stable vacuum at $\phi=0$. Expanding the potential at small $\g$ and diagonalising the mass matrix one obtains the expansion \eqref{eq0_1}. There are two classes of bosonic affine Toda theories: untwisted and twisted models. In untwisted theories, $\alpha_0$ corresponds to the negative of the highest root of $\mathfrak{g}$ and in these models all nonzero $3$-point couplings satisfy the following fusing rule~\cite{Dorey:1990xa}
\begin{equation}
\label{eq:coupling_fusing_rule}
    C^{(3)}_{abc} \ne 0  \iff \exists \ \alpha \in \Gamma_a, \, \beta \in \Gamma_b,\, \gamma \in \Gamma_c \  \text{such that} \ \alpha+\beta+\gamma=0\,,
\end{equation}
where $\Gamma_a$, $\Gamma_b$ and $\Gamma_c$ are Coxeter orbits, as defined in~\cite{Dorey:1990xa}. The nonvanishing couplings satisfy then the area rule~\cite{Braden:1989bu,Fring:1991me}
\begin{equation}
\label{eq:coupling_area_rule2}
    C^{(3)}_{abc} = \frac{4 \g}{\sqrt{h}}\sigma_{abc} \ \Delta_{abc} \,,
\end{equation}
where $h$ is the Coxeter number of $\mathfrak{g}$
and $\sigma_{abc}$, up to a possible sign, are the structure constants of the underlying algebra~\cite{Fring:1991me}.
Famous examples of these models are the simply-laced theories, constructed from simply connected Dyknin diagrams, for which the couplings can be normalised so that $\sigma_{abc}= \pm 1$. The second class is composed of twisted models. The masses and couplings of each one of these theories are a subsector of the masses and couplings of a simply-laced model~\cite{Braden:1989bu}, from which the twisted model can be obtained through the so-called folding procedure. Classically these theories can then be described as truncations of simply-laced theories (we remand to appendices~\ref{sec:untwisted-theories} and~\ref{sec:twisted-theories} for more details). In the rest of this section, we will consider nonsimply-laced affine Toda theories, comprising both twisted models and untwisted models based on non-simply-laced Dynkin diagrams.

\subsection{One-loop S-matrices}
\label{subsec_one_loop_Smat}

We now use expression~\eqref{eq:result_S_mat_eq_m_ren_3} to write the one-loop S-matrices of nonsimply-laced affine Toda theories. To perform the computation we use that in all affine Toda models~\cite{Dorey:2021hub}
\begin{equation}
\label{eq:ainfinity_Toda}
\ai= - i \frac{\g^2}{\mu^2 h}\,,
\end{equation}
and
\begin{equation}
\label{eq:Mcollin_Toda}
\frac{M^{(0)}_{jj}(0)}{m^2_j}= -i \frac{2 \g^2}{h} \alpha^2_j\quad \forall \ j \in \{1,\, \dots, \, r\} \,,
\end{equation}
with $\alpha_j$ being the simple root of the Dynkin diagram associated with the particle of type $j$.
Let $\mathbb{M}^2$ be the squared mass matrix, defined through~\eqref{eq:lagrangian-toda} by
\begin{equation}
\text{Tr} \bigl(\mathbb{M}^2 \bigl) =\sum^r_{e=1} m_e^2= \mu^2 \sum^r_{i=0} n_i \alpha^2_i \,.
\end{equation}
Using these relations we find that the one-loop S-matrix of any affine Toda model is a function of the underlying Lie algebra data and the tree-level S-matrices of the model. Indeed, using~\eqref{eq:ainfinity_Toda} and~\eqref{eq:Mcollin_Toda}, then~\eqref{eq:result_S_mat_eq_m_ren_3} can be written as
\begin{equation}
\label{eq:result_S_mat_eq_m_ren_4}
\begin{split}
S^{(1)}_{ab}(\theta_{pp'})&=\frac{\bigl(S^{(0)}_{ab}(\theta_{p p'}) \bigl)^2}{2}+\frac{\text{Tr} \bigl(\mathbb{M}^2 \bigl)}{8 \pi} \frac{\g^2}{\mu^2 h}  \theta_{p p'} \frac{\partial}{\partial \theta_{p p'}} S^{(0)}_{ab}(\theta_{p p'})  \\
&-\frac{1}{4n \pi^2}  \sum_{e=1}^r \frac{\partial}{\partial \theta_{p p'}} \oint_{\ip_n} d\theta_k \theta_k S^{(0)}_{ea} (\theta_{k p} ) S^{(0)}_{e b} (\theta_{k p'})\\
&-\frac{i}{8} \frac{\g^2}{h}  \left( \alpha^2_a - \alpha^2_b \right) \frac{\partial}{\partial \theta_{p p'}} S^{(0)}_{ab}(\theta_{p p'}) \,.
\end{split}
\end{equation}

Plugging the tree-level S-matrices of affine Toda theories into~\eqref{eq:result_S_mat_eq_m_ren_4}, we computed the one-loop S-matrices for all nonsimply-laced models and found exact agreement with the results bootstrapped in~\cite{Delius:1991kt,Corrigan:1993xh}. This provides a strong confirmation of the exactness of formula~\eqref{eq:result_S_mat_eq_m_ren_3} obtained from perturbation theory and a systematic check that the results bootstrapped in the past for nonsimply-laced models are correct. Because these checks are just a direct application of \eqref{eq:result_S_mat_eq_m_ren_4} to all nonsimply-laced models, instead of showing the checks for all these instances we will illustrate the overall logic in a simple example below.

\paragraph{The dual-class $(g^{(1)}_{2}, d^{(3)}_4)$.} 
The $g^{(1)}_{2}$ and $d^{(3)}_4$ affine Toda theories are constructed from Dynkin diagrams that map one into the other under inversion of the root lengths.
They are among the simplest examples of nonsimply-laced affine Toda models since have only two particles in the spectrum with \emph{classical} mass ratios given by
\begin{equation}
\label{eq:mass_ratio_G2}
\begin{split}
    & g_2^{(1)} :  \frac{m_2^2}{m_1^2} = 3\, , \\
    & d_4^{(3)} : \frac{m_2^2}{m_1^2} = 2+\sqrt{3} \,.
\end{split}
\end{equation}
As aforementioned, an exact quantum S-matrix for this class was proposed in~\cite{Delius:1991kt}, interpolating at the strong and weak coupling between the classical S-matrices obtained from the Lagrangians of the two elements of the dual pair.  In the proposal of~\cite{Delius:1991kt} there is a one-parameter mass ratio given by
\begin{equation}
\label{eq:exact-mass-ratio}
    \frac{\hat{m}_2^2}{\hat{m}_1^2} = \frac{\sin^2 (2\pi/H)}{\sin^2 (\pi/H)},
\end{equation}
where $H=6+3B$ and $B\in[0,2]$. Note here the use of $\hat{m}_k^2$ to denote the \emph{physical} masses. Interestingly $H$ plays the role of a floating Coxeter number since it interpolates between the Coxeter numbers of $g_2^{(1)}$ and $d_4^{(3)}$, which are $h=6$ and $h=12$ respectively and are realised at the boundary values $B=0$ and $B=2$. We can also see this interpolation between the elements of the dual pair from the spectrum, indeed for $B\rightarrow 0$ and $B\rightarrow 2$ we approximate the $g^{(1)}_{2}$ and $d^{(3)}_4$ spectrum, respectively. The parameter $B$ is a function of the Lagrangian coupling constant $g$ and it was conjectured to be\footnote{The same parameter also introduces the coupling dependence in the exact S-matrices of simply-laced theories~\cite{Braden:1989bu}.}
\begin{equation}
    B = \frac{g^2}{2\pi} \frac{1}{1+g^2 /4\pi} \, .
\end{equation}
Then we have indeed a weak-strong coupling interpolation between these two models. By computing the mass shifts $\delta m_j^2$ for these two models (see appendix \ref{app:mass-shifts}) and by using the definition \eqref{eq:ren_masses_couplings} for the physical masses we find the following physical mass ratios  
\begin{equation}
\begin{split}
    & g_2^{(1)} : \frac{\hat{m}_2^2}{\hat{m}_1^2} = 3+\frac{g^2}{4
   \sqrt{3}}+O\left(g^3\right), \\
    & d_4^{(3)} : \frac{\hat{m}_2^2}{\hat{m}_1^2} = 2+\sqrt{3} -\frac{
   g^2}{48}+O\left(g^3\right) .
\end{split}
\end{equation}
These exactly coincide with the expansion of \eqref{eq:exact-mass-ratio} up to order $O(g^2)$, thus confirming its validity from a perturbation theory approach\footnote{Note that for the $d_4^{(3)}$ mass ratio we expanded around $2-B$ and then $g\rightarrow 0$ which is equivalent to the $g\rightarrow \infty$ expansion after redefining the coupling.}.

Similarly, the S-matrix proposed in~\cite{Delius:1991kt} for the dual pair interpolates through $H$ at strong and weak coupling the tree-level S-matrices of $g_2^{(1)}$ and $d_4^{(3)}$. The quantum S-matrix is defined in terms of the following building blocks
\begin{equation}
\label{eq:gen-blocks1}
\begin{split}
& (x)=\frac{\sinh \left( \frac{\theta}{2} + \frac{i \pi x}{2 H} \right)}{\sinh \left( \frac{\theta}{2} - \frac{i \pi x}{2 H} \right)}, \\
& \{ x \}_{\nu}= \frac{(x-\nu B -1) (x+\nu B+1)}{(x+\nu B-1+B) (x-\nu B+1-B)},
\end{split}
\end{equation}
which are modifications of the building blocks of the simply-laced theories introduced in~\cite{Braden:1989bu}, with the latter being recovered in the case $\nu = 0$. The set of S-matrices for this dual pair is given by
\begin{equation}
\label{eq:g2d4Smatrix}
    \begin{split}
        S_{11} (\theta) & = \{1\}_{0} \{ H/2 \}_{1/2} \{ H-1 \}_{0} ,\\
        S_{12} (\theta) & = \{H/3\}_{1} \{ 2H/3 \}_{1} ,\\
        S_{22} (\theta) & = \left\{\frac{H}{3}-1\right\}_{1} \left\{\frac{H}{3}+1\right\}_{1} \left\{\frac{2H}{3}-1\right\}_{1} \left\{\frac{2H}{3}+1\right\}_{1} .
    \end{split}
\end{equation}
The logic for our check is then the following.
First, we verified at both endpoints of $H$ that the S-matrices above correspond to the tree-level S-matrices obtained from the Lagrangians~\eqref{eq:lagrangian-toda} for these models. Then, by inserting these tree-level S-matrices into the one-loop formula~\eqref{eq:result_S_mat_eq_m_ren_4}, we precisely reproduced the $O(g^4)$ contribution obtained from~\eqref{eq:g2d4Smatrix} for both $g_2^{(1)}$ and $d_4^{(3)}$. We did a similar analysis for all the remaining nonsimply-laced Toda models and verified that the conjectured S-matrices of~\cite{Delius:1991kt,Corrigan:1993xh} describe these theories up to one loop with all of them being integrable up to this loop order.

\subsection{Landau poles in inelastic processes}
\label{sec_Landau_poles_inel_processes}

While Landau poles in inelastic processes always cancel against the expansion of the tree-level amplitude around the classical mass values (see~\eqref{eq:cancel_double_poles_expanding_tree_amplitude}) it is anyway interesting to see how formula~\eqref{ab_to_cd_one_loop_inelastic_amp_derivatives} correctly reproduces the correct behaviour of one-loop Landau diagrams of inelastic processes.
In this section we explicitly check the values of the second-order poles predicted by~\eqref{ab_to_cd_one_loop_inelastic_amp_derivatives} against Landau diagrams in the $g_2^{(1)}$ model. Many other examples have been similarly considered; all of them indicate that the singular behaviour of one-loop Landau diagrams is always encoded in the values of the mass shifts, according to formula~\eqref{ab_to_cd_one_loop_inelastic_amp_derivatives}.

As previously described, the $g_2^{(1)}$ affine Toda model has two particles in the spectrum. We focus on the following inelastic process
\begin{equation}
\label{eq:22_to_11_G2}
2(p)+2(p') \to 1(q)+1(q'),
\end{equation}
This process has Feynman diagrams with simultaneous singularities at $s=m_2^2$ and $u=m_1^2$, in agreement with the flipping rule.
\begin{figure}
\begin{center}
\begin{tikzpicture}
\tikzmath{\y=1.5;}
\tikzmath{\x=1;}

%Diagram 1
\draw[directed] (0*\y,0*\y) -- (-1.5*\y,0.866025*\y);
\draw[directed] (-1.5*\y,0.866025*\y) -- (0*\y,2*0.866025*\y);
\draw[directed] (0*\y,0*\y) -- (0.5*\y,0.866025*\y);
\draw[directed] (0.5*\y,0.866025*\y) -- (0*\y,2*0.866025*\y);
\draw[] (0*\y,0*\y) -- (0*\y,2*0.866025*\y);
\filldraw[black] (-1*\y,0.1*\y)  node[anchor=west] {\scriptsize{$2(p')$}};
\filldraw[black] (-1*\y,1.6*\y)  node[anchor=west] {\scriptsize{$2(p)$}};
\filldraw[black] (-0.2*\y,1*\y)  node[anchor=west] {\scriptsize{$2$}};
\filldraw[black] (0.2*\y,0.1*\y)  node[anchor=west] {\scriptsize{$1(q)$}};
\filldraw[black] (0.2*\y,1.6*\y)  node[anchor=west] {\scriptsize{$1(q')$}};
\draw[directed] (-2+0.5*\x,4.2+0.866025*\x) -- (-2+1*\x,4.2+0*\x);
\draw[directed] (-2+0.5*\x,4.2-0.866025*\x) -- (-2+1*\x,4.2+0*\x);
\draw[] (-2+1*\x,4.2+0*\x) -- (-2+2*\x,4.2+0*\x);
\draw[directed] (-2+2*\x,4.2+0*\x) -- (-2+2*\x+0.866025*\x,4.2+0*\x+0.5*\x);
\draw[directed] (-2+2*\x,4.2+0*\x) -- (-2+2*\x+0.866025*\x,4.2+0*\x-0.5*\x);
\filldraw[black] (-2+0.5*\x,4.2+0.866025*\x)  node[anchor=west] {\scriptsize{$2(p)$}};
\filldraw[black] (-2+0.5*\x,4.2-0.866025*\x)  node[anchor=west] {\scriptsize{$2(p')$}};
\filldraw[black] (-2+1.3*\x,4.2+0.2*\x)  node[anchor=west] {\scriptsize{$2$}};
\filldraw[black] (-2+2.5*\x,4.2-0.866025*\x)  node[anchor=west] {\scriptsize{$1(q)$}};
\filldraw[black] (-2+2.5*\x,4.2+0.866025*\x)  node[anchor=west] {\scriptsize{$1(q')$}};

%Diagram 2
\draw[directed] (0*\y+6,0*\y) -- (-1.5*\y+6,0.866025*\y);
\draw[directed] (-1.5*\y+6,0.866025*\y) -- (0*\y+6,2*0.866025*\y);
\draw[directed] (+6,0.0*\y) -- (-0.5*\y+6,0.866025*\y);
\draw[directed] (-0.5*\y+6,0.866025*\y) -- (0*\y+6,2*0.866025*\y);
\draw[] (-0.5*\y+6,0.866025*\y) -- (-1.5*\y+6,0.866025*\y);
\filldraw[black] (-1*\y+6,0.1*\y)  node[anchor=west] {\scriptsize{$2(p')$}};
\filldraw[black] (-1*\y+6,1.6*\y)  node[anchor=west] {\scriptsize{$2(p)$}};
\filldraw[black] (-0.9*\y+6,1*\y)  node[anchor=west] {\scriptsize{$1$}};
\filldraw[black] (-0.2*\y+6,0.4*\y)  node[anchor=west] {\scriptsize{$1(q')$}};
\filldraw[black] (-0.2*\y+6,1.3*\y)  node[anchor=west] {\scriptsize{$1(q)$}};
\draw[directed] (-2+0.5*\x+5,4.5+0.866025*\x+0.5*\x) -- (-2+1*\x+5,4.5+0*\x+0.5*\x);
\draw[directed] (-2+0.5*\x+5,4.5-0.866025*\x-1*\x) -- (-2+1*\x+5,4.5+0*\x-1*\x);
\draw[] (-2+1*\x+5,4.5+0.5*\x) -- (-2+1*\x+5,4.5-1*\x);
\draw[directed] (-2+1*\x+5,4.5-1*\x) -- (-2+1*\x+0.866025*\x+5,4.5+0*\x-0.5*\x);
\draw[directed] (-2+1*\x+5,4.5+0.5*\x) -- (-2+1*\x+0.866025*\x+5,4.5);
\filldraw[black] (-2+0.3*\x+4.5,4.2+1.2*\x)  node[anchor=west] {\scriptsize{$2(p)$}};
\filldraw[black] (-2+0.3*\x+4.5,4.2-1.2*\x)  node[anchor=west] {\scriptsize{$2(p')$}};
\filldraw[black] (-2+1.1*\x+4.5,4.2+0.1*\x)  node[anchor=west] {\scriptsize{$1$}};
\filldraw[black] (-2+2*\x+4.5,4.2-0.6*\x)  node[anchor=west] {\scriptsize{$1(q')$}};
\filldraw[black] (-2+2*\x+4.5,4.2+0.6*\x)  node[anchor=west] {\scriptsize{$1(q)$}};

\end{tikzpicture}
\caption{Pair of simultaneously singular on-shell diagrams in the $g^{(1)}_2$ affine Toda theory (on the bottom) and associated Feynman diagrams (on the top) for the $2+2\rightarrow 1+1$ process.}
\label{Pair_of_singular_diagram_in_G2}
\end{center}
\end{figure}
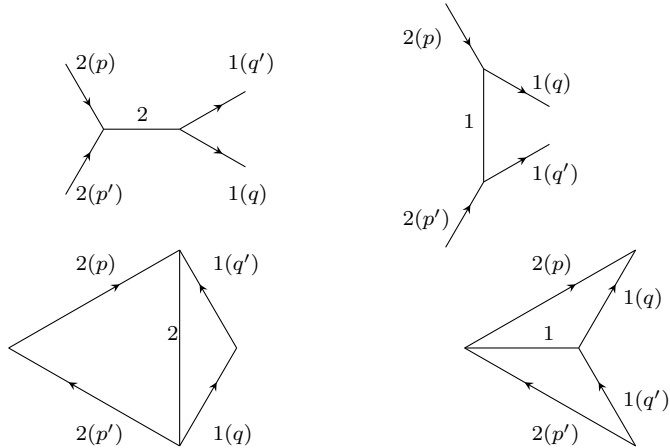
The pair of simultaneously on-shell diagrams is shown in figure~\ref{Pair_of_singular_diagram_in_G2} and their sum has vanishing residue at the pole in the tree-level amplitude evaluated at the classical values of the masses. Then to have the absence of double poles in one-loop diagrams the constraint~\eqref{eq:residue_pole_oneloop_generic} must hold; in this case (since there is no on-shell particle in the $t$-channel) the constraint becomes
\begin{equation}
    \frac{\sigma_{222} \sigma_{112}}{\Delta_{222} \Delta_{112}} \mathcal{F}_s^{(2)}(m_2^2) + \frac{\sigma^2_{112}}{\Delta^2_{112} } \mathcal{F}_u^{(2)}(m^2_2, m_1^2) = 0 \,,
\end{equation}
where we used the area rule for the couplings written in~\eqref{eq:coupling_area_rule2}.
After substituting the values of $\sigma_{abc}$ defined in~\eqref{eq:g2-couplings} for this model,
the constraint above yields the following relation for the mass corrections
\begin{equation}
\label{eq:mass_correction_G2_constraint}
    \frac{\delta m_2^2}{\delta m_1^2} =  3 \, .
\end{equation}
Note that the constraint for the absence of double poles in the one-loop amplitude associated with the process~\eqref{eq:22_to_11_G2} is that the radiative corrections do not affect the mass ratio in the first line of~\eqref{eq:mass_ratio_G2}. 
Since~\eqref{eq:mass_correction_G2_constraint} is incompatible with the actual radiative corrections of the masses, as shown from~\eqref{eq:312245}, the one-loop inelastic amplitude features a double pole. Let us compute this pole.

Using the couplings $\sigma_{abc}$ and the correct radiative corrections of the masses given in \eqref{eq:g2-couplings} and \eqref{eq:312245}, respectively, we observe the following nonvanishing double pole in the one-loop amplitude 
\begin{equation}
\label{eq:839894}
\begin{split}
    M^{(1)}_{22\rightarrow11} & \sim \frac{-i}{(s-m_2^2)^2} \left(\frac{4 \g}{\sqrt{h}} \right)^2 \Delta^2_{222} \Delta^2_{112} \left(  \frac{\sigma_{222} \sigma_{112}}{\Delta_{222} \Delta_{112}}\ \mathcal{F}_s^{(2)}(m_2^2) +  \frac{\sigma^2_{112}}{\Delta^2_{112}} \mathcal{F}_u^{(2)}(m^2_2, m_1^2)\right), \\ 
    & \sim \frac{-i\  \sqrt{3}\ g^4 \mu^6}{(s-m_2^2)^2}  .
\end{split}
\end{equation}
This double-pole coefficient (obtained above from the universal formula~\eqref{ab_to_cd_one_loop_inelastic_amp_derivatives}) can be reproduced by a study of Landau diagrams.
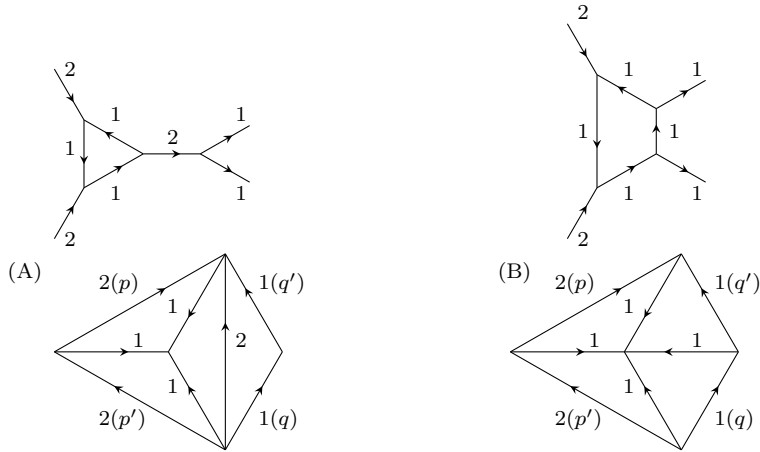
\begin{figure}
\begin{center}
\begin{tikzpicture}
\tikzmath{\y=1.5;}
\tikzmath{\x=1;}

%Diagram 1
\filldraw[black] (-0.5*\y,0*\y+0.7*\y)  node[anchor=west] {\scriptsize{(A)}};
\draw[directed] (1.5*\y,-0.866025*\y) -- (0*\y,0*\y);
\draw[directed] (0*\y,0*\y) -- (1.5*\y,0.866025*\y);
\draw[directed] (1.5*\y,-0.866025*\y) -- (1.5*\y,0.866025*\y);
\draw[directed] (0*\y,0*\y) -- (1*\y,0*\y);
\draw[directed] (1.5*\y,-0.866025*\y) -- (1*\y,0*\y);
\draw[directed] (1.5*\y,0.866025*\y) -- (1*\y,0*\y);
\draw[directed] (1.5*\y,-0.866025*\y) -- (2*\y,0*\y);
\draw[directed] (2*\y,0*\y) -- (1.5*\y,0.866025*\y);
\filldraw[black] (0.3*\y,0.6*\y)  node[anchor=west] {\scriptsize{$2(p)$}};
\filldraw[black] (0.3*\y,-0.6*\y)  node[anchor=west] {\scriptsize{$2(p')$}};
\filldraw[black] (1.7*\y,0.6*\y)  node[anchor=west] {\scriptsize{$1(q')$}};
\filldraw[black] (1.7*\y,-0.6*\y)  node[anchor=west] {\scriptsize{$1(q)$}};
\filldraw[black] (0.6*\y,0.1*\y)  node[anchor=west] {\scriptsize{$1$}};
\filldraw[black] (0.9*\y,0.4*\y)  node[anchor=west] {\scriptsize{$1$}};
\filldraw[black] (0.9*\y,-0.3*\y)  node[anchor=west] {\scriptsize{$1$}};
\filldraw[black] (1.5*\y,0.1*\y)  node[anchor=west] {\scriptsize{$2$}};
\draw[directed] (0*\y,0*\y+1*\y) -- (0.26*\y, 0.45*\y+1*\y);
\draw[directed] (0*\y,0.9*\y+1*\y+0.6*\y) -- (0.26*\y, 0.45*\y+1*\y+0.6*\y);
\draw[directed] (0.26*\y, 0.45*\y+1*\y+0.6*\y) -- (0.26*\y, 0.45*\y+1*\y);
\draw[directed] (0.26*\y, 0.45*\y+1*\y) -- (0.26*\y+0.52*\y, 0.45*\y+1*\y+0.3*\y);
\draw[directed] (0.26*\y+0.52*\y, 0.45*\y+1*\y+0.3*\y) -- (0.26*\y, 0.45*\y+1*\y+0.6*\y);
\draw[directed] (0.26*\y+0.52*\y, 0.45*\y+1*\y+0.3*\y) -- (0.76*\y+0.52*\y, 0.45*\y+1*\y+0.3*\y);
\draw[directed] (0.76*\y+0.52*\y, 0.45*\y+1*\y+0.3*\y) -- (0.76*\y+0.52*\y+0.433*\y, 0.45*\y+1*\y+0.3*\y+0.25*\y);
\draw[directed] (0.76*\y+0.52*\y, 0.45*\y+1*\y+0.3*\y) -- (0.76*\y+0.52*\y+0.433*\y, 0.45*\y+1*\y+0.3*\y-0.25*\y);
\filldraw[black] (0*\y,0*\y+1*\y)  node[anchor=west] {\scriptsize{$2$}};
\filldraw[black] (0*\y,0*\y+2.5*\y)  node[anchor=west] {\scriptsize{$2$}};
\filldraw[black] (0.9*\y,0*\y+1.9*\y)  node[anchor=west] {\scriptsize{$2$}};
\filldraw[black] (1.5*\y,0*\y+2.1*\y)  node[anchor=west] {\scriptsize{$1$}};
\filldraw[black] (1.5*\y,0*\y+1.4*\y)  node[anchor=west] {\scriptsize{$1$}};
\filldraw[black] (0.4*\y,0*\y+1.4*\y)  node[anchor=west] {\scriptsize{$1$}};
\filldraw[black] (0.4*\y,0*\y+2.1*\y)  node[anchor=west] {\scriptsize{$1$}};
\filldraw[black] (0*\y,0*\y+1.8*\y)  node[anchor=west] {\scriptsize{$1$}};

%Diagram 2
\filldraw[black] (-0.5*\y+4.3*\y,0*\y+0.7*\y)  node[anchor=west] {\scriptsize{(B)}};
\draw[directed] (1.5*\y+4*\y,-0.866025*\y) -- (0*\y+4*\y,0*\y);
\draw[directed] (0*\y+4*\y,0*\y) -- (1.5*\y+4*\y,0.866025*\y);
\draw[directed] (0*\y+4*\y,0*\y) -- (1*\y+4*\y,0*\y);
\draw[directed] (2*\y+4*\y,0*\y) -- (1*\y+4*\y,0*\y);
\draw[directed] (1.5*\y+4*\y,-0.866025*\y) -- (1*\y+4*\y,0*\y);
\draw[directed] (1.5*\y+4*\y,0.866025*\y) -- (1*\y+4*\y,0*\y);
\draw[directed] (1.5*\y+4*\y,-0.866025*\y) -- (2*\y+4*\y,0*\y);
\draw[directed] (2*\y+4*\y,0*\y) -- (1.5*\y+4*\y,0.866025*\y);
\filldraw[black] (0.3*\y+4*\y,0.6*\y)  node[anchor=west] {\scriptsize{$2(p)$}};
\filldraw[black] (0.3*\y+4*\y,-0.6*\y)  node[anchor=west] {\scriptsize{$2(p')$}};
\filldraw[black] (1.7*\y+4*\y,0.6*\y)  node[anchor=west] {\scriptsize{$1(q')$}};
\filldraw[black] (1.7*\y+4*\y,-0.6*\y)  node[anchor=west] {\scriptsize{$1(q)$}};
\filldraw[black] (0.6*\y+4*\y,0.1*\y)  node[anchor=west] {\scriptsize{$1$}};
\filldraw[black] (0.9*\y+4*\y,0.4*\y)  node[anchor=west] {\scriptsize{$1$}};
\filldraw[black] (0.9*\y+4*\y,-0.3*\y)  node[anchor=west] {\scriptsize{$1$}};
\filldraw[black] (1.5*\y+4*\y,0.1*\y)  node[anchor=west] {\scriptsize{$1$}};
\draw[directed] (0*\y+4.5*\y,0*\y+1*\y) -- (0.26*\y+4.5*\y, 0.45*\y+1*\y);
\draw[directed] (0*\y+4.5*\y,0.9*\y+1*\y+1*\y) -- (0.26*\y+4.5*\y, 0.85*\y+1*\y+0.6*\y);
\draw[directed] (0.26*\y+4.5*\y, 0.85*\y+1*\y+0.6*\y) -- (0.26*\y+4.5*\y, 0.45*\y+1*\y);
\draw[directed] (0.26*\y+4.5*\y, 0.45*\y+1*\y) -- (0.26*\y+0.52*\y+4.5*\y, 0.45*\y+1*\y+0.3*\y);
\draw[directed] (0.26*\y+0.52*\y+4.5*\y, 0.85*\y+1*\y+0.3*\y) -- (0.26*\y+4.5*\y, 0.65*\y+1*\y+0.8*\y);
\draw[directed] (0.78*\y+4.5*\y, 1.75*\y) -- (0.78*\y+4.5*\y, 2.15*\y);
\draw[directed] (0.78*\y+4.5*\y, 2.15*\y) -- (0.78*\y+4.5*\y+0.433*\y, 2.15*\y+0.25*\y);
\draw[directed] (0.78*\y+4.5*\y, 1.75*\y) -- (0.78*\y+4.5*\y+0.433*\y, 1.75*\y-0.25*\y);
\filldraw[black] (0*\y+4.5*\y,0*\y+1*\y)  node[anchor=west] {\scriptsize{$2$}};
\filldraw[black] (0*\y+4.5*\y,0*\y+3*\y)  node[anchor=west] {\scriptsize{$2$}};
\filldraw[black] (0.8*\y+4.5*\y,0*\y+1.95*\y)  node[anchor=west] {\scriptsize{$1$}};
\filldraw[black] (1*\y+4.5*\y,0*\y+2.5*\y)  node[anchor=west] {\scriptsize{$1$}};
\filldraw[black] (1*\y+4.5*\y,0*\y+1.4*\y)  node[anchor=west] {\scriptsize{$1$}};
\filldraw[black] (0*\y+4.5*\y,0*\y+1.95*\y)  node[anchor=west] {\scriptsize{$1$}};
\filldraw[black] (0.4*\y+4.5*\y,0*\y+2.5*\y)  node[anchor=west] {\scriptsize{$1$}};
\filldraw[black] (0.4*\y+4.5*\y,-0.1*\y+1.5*\y)  node[anchor=west] {\scriptsize{$1$}};

\end{tikzpicture}
\caption{On-shell Landau diagrams (on the top) contributing to the scattering $2+2 \to 1+1$ at one-loop in the $g^{(1)}_2$ affine Toda theory and their duals (on the bottom).}
\label{Landau_diagrams_G2}
\end{center}
\end{figure}
We notice that at $s=m_2^2$ there are two Landau diagrams contributing to the double pole, with these depicted in figure~\ref{Landau_diagrams_G2}. If our analysis is correct then we should obtain the same residue by summing the two diagrams.
Using either the approach of~\cite{Braden:1990wx} or the cutting technique discussed in~\cite{Dorey:2022fvs,Dorey:2023cuq} we find out the following values for the Landau diagrams at the singularity
\begin{equation}
\begin{split}
D^{\text{(A)}}& =i \sqrt{3} \frac{\g^4 \mu^6}{(s- m_2^2)}\,,\\
D^{\text{(B)}}&=-2i \sqrt{3} \frac{ \g^4 \mu^6}{(s- m_2^2)}\,,
\end{split}
\end{equation}
whose sum exactly reproduces~\eqref{eq:839894}; this check provides evidence of the validity of formula~\eqref{ab_to_cd_one_loop_inelastic_amp_derivatives}.
We remark one more time that the presence of this second-order pole is not a problem for the one-loop integrability of the model since it is cancelled against the expansion of the tree-level amplitude around the classical masses.

The study of Landau singularities in elastic processes is more subtle and for non-simply laced theories can be performed following two different routes.
The reason is the following. Suppose to know the exact non-perturbative S-matrix of the model (which can be bootstrapped through unitarity, analyticity, crossing symmetry etc). This S-matrix has some poles and a certain dependence on the coupling $g$. We can either expand in $g$ and then evaluate the residues of the poles (in perturbation theory this corresponds to a loop-by-loop expansions using the classical masses in the Landau diagrams) or evaluate the residues at the poles and expand in small $g$ afterwards (in perturbation theory this corresponds to use the renormalized values of the masses in Landau diagrams). The small $g$ limit does not have to commute with the Laurent expansion around a pole and if the mass ratios are affected by quantum corrections the two limiting procedures can generate different answers. In summary, we have the following two routes and we stress that the Landau diagrams of this paper have been computed following the first route.
\begin{enumerate}
    \item[(1)] For a given pole one can first expand the exact S-matrix at small $\g$ and then evaluate the one-loop order of the expansion at the pole. The residue obtained in this way is fixed and can be matched with the one obtained by summing over one-loop Landau diagrams (this is similar to what we did in this section for inelastic processes).  Following this route, the masses of particles appearing in one-loop Landau diagrams must be fixed to their classical values. This corresponds to an honest loop expansion of the amplitude, in which we first expand in $\g$ (we must expand also the renormalized masses in $\g$ as $\hat{m}^2_j=m^2_j+\delta m^2_j+\dots$) and then expand around the pole positions predicted by the classical masses.

    \item[(2)] The second route is to first expand the exact S-matrix at the pole position (which is a function of the coupling $\g$ predicted by the renormalized masses that we leave unexpanded) and only after, eventually, to perform the expansion of the residue at small coupling $\g$. This procedure generates in general a pole of different order compared to the one expected from the expansion done following route (1).
\end{enumerate} 
While it is quite simple to confirm the expansion done following root (1) through perturbation theory, it is difficult to confirm the expansion done in the second way through perturbative computations.
This is because we are expanding around a pole which has an exact dependence on the coupling and we should sum over infinitely many Feynman diagrams (contributing at all higher-loop orders) to match the residue obtained from the exact S-matrix. In~\cite{Delius:1991kt,Corrigan:1993xh} it was discussed a generalised Coleman-Thun mechanism for the generation of the residues following this second route, even though the exact computation of these residues was not performed in most of the cases. More recently, a way for computing non-perturbatively double poles in elastic amplitudes was investigated in~\cite{Correia:2022dcu} and tested on the $E^{(1)}_8$ simply-laced model\footnote{We thank the anonymous referee for having pointed this out to us.}, for which the mass ratios do not receive one-loop corrections and the pole positions are expected to be coupling independent at all loop orders. It would be interesting to see if the same method can be extended to non-simply laced theories, where pole positions depend non-perturbatively on the coupling.

We also stress that formula~\eqref{ab_to_cd_one_loop_inelastic_amp_derivatives} implies that double poles in a one-loop inelastic amplitude are always associated with the propagation of at least one bound state in the associated tree-level amplitude. Sums of Landau diagrams in which no bound states propagate at the tree level (which means all diagrams are of the type on the RHS of figure~\ref{Landau_diagrams_G2} and there is no diagram of the type on the LHS) must cancel. It is a known fact that these diagrams exist in general and even-order poles in elastic amplitudes of simply-laced affine Toda theories are explained through them (see, e.g. \cite{Dorey:2022fvs}). However, for inelastic processes, their sums must always vanish; this is a consequence of equation~\eqref{ab_to_cd_one_loop_inelastic_amp_derivatives}.

\section{Conclusion}
\label{sec:Conclusions}

In this paper, we generalised the results of~\cite{Polvara:2023vnx,Fabri:2024qgd} to models with mass ratios which are affected by one-loop quantum corrections. We showed that any theory with a Lagrangian of type~\eqref{eq0_1} which is purely elastic at the tree level is also purely elastic at one-loop; for this to be true we needed to define the renormalized masses of the theory ($\{\hat{m}_a \}^r_{a=1}$) to be shifted with respect to the classical ones ($\{m_a \}^r_{a=1}$) by radiative corrections carried by one-loop bubble diagrams. Once the physical renormalized masses are defined in this way all one-loop inelastic processes vanish and the one-loop exact S-matrices are functions of the tree-level S-matrices through the universal formula~\eqref{eq:result_S_mat_eq_m_ren_3}. We checked such a formula systematically on the class of nonsimply-laced affine Toda theories and we were able to reproduce the one-loop expansion of the exact S-matrices of these models known in the literature~\cite{Delius:1991kt, Corrigan:1993xh}.
As a side result, we also showed how the corrections to the masses (i.e. the set $\{ \delta m^2_a \}^r_{a=1}$) carry total information on the Landau singularities of one-loop inelastic amplitudes. For some simple examples, we computed second-order poles both through equation~\eqref{ab_to_cd_one_loop_inelastic_amp_derivatives} (which uses mass corrections) and the standard Landau analysis. In all the considered cases we obtained that the residues at the Landau poles are in agreement with the expectation of~\eqref{ab_to_cd_one_loop_inelastic_amp_derivatives}. These residues vanish only if the mass ratios do not receive quantum corrections while in all the other cases the singularities cancel in the full amplitude against the expansion of the tree-level amplitude around the classical values of the masses, as shown in~\eqref{eq:cancel_double_poles_expanding_tree_amplitude}.

It would be interesting to try to simplify formula~\eqref{eq:result_S_mat_eq_m_ren_3} further and search for a connection with similar results obtained through unitarity-cuts methods developed in the past~\cite{Bianchi:2014rfa,Bianchi:2013nra}. While the unitarity cut methods used in~\cite{Bianchi:2014rfa,Bianchi:2013nra} fail to reproduce the correct one-loop S-matrices of affine Toda models, it would be anyway interesting to determine under which conditions those formulas are valid. 
It is remarkable indeed that the formulas proposed in~\cite{Bianchi:2014rfa,Bianchi:2013nra} have been able to reproduce the correct worldsheet S-matrices of superstrings propagating in AdS$_5 \times$ S$^5$ and AdS$_3 \times$ S$^3 \times$ T$^4$, while they fail for the cases under discussion~\cite{Fabri:2024qgd}.

Another natural direction is to use the universal expression for the tree-level S-matrices of affine Toda theories (these S-matrices can be derived from perturbation theory and can be completely written in terms of roots of their underlying Lie algebras~\cite{Dorey:2021hub}), to generate a universal expression for the one-loop S-matrices of these models. So far universal expressions for these one-loop S-matrices have been derived from perturbation theory only in simply-laced theories~\cite{Fabri:2024qgd} while the study performed in section~\ref{subsec_one_loop_Smat} on non-simply-laced models remains based on case-by-case analysis. 
A more universal understanding of formula~\eqref{eq:result_S_mat_eq_m_ren_4} would probably lead to the expansion of the S-matrices proposed in~\cite{Oota:1997un} using a $q$-deformed Coxeter element. 

Last but not least it would be very interesting to investigate the generalisation of formula~\eqref{eq:result_S_mat_eq_m_ren_3} to higher-loop and to more general classes of tree-level integrable models. 

In this paper, we wrote one-loop amplitudes in terms of tree-level amplitudes using the method
proposed in~\cite{Polvara:2023vnx,Fabri:2024qgd}.
The fact that one-loop amplitudes can be written entirely in terms of tree-level amplitudes comes from the factorisability of the tree-level S-matrices, which allows us to write the integrand in the second line of~\eqref{eq:splitting_ono_loop_amplitude_into_double_and_single_cuts} (when $\{a,b \}= \{c, d\}$) as a product of two-body amplitudes.
Using the same technique on a two-loop amplitude $M_{ab \to ab}^{(2)}$  we would generate integrals of four-to-four tree-level amplitudes of type $\sum^r_{ef=1} \int d \theta_e d \theta_f M_{ab e f \to ab e f}^{(0)}$ in which the integral is performed over the two-dimensional space of the rapidities $\theta_e$ and $\theta_f$. Since we are assuming the theory to be tree-level integrable it must be possible to write the integrand $M_{ab e f \to ab e f}^{(0)}$ as a product of two-body tree-level amplitude (this needs to be done taking care of removing collinear singularities through counterterms, as done in~\cite{Fabri:2024qgd} for the one-loop case). It would be very interesting to see if there exists a closed recursion formula to generate higher-loop S-matrices of tree-level integrable theories in terms of tree amplitudes. 
We hope to return to some of these problems in the future.

\section*{Acknowledgments}
DP thanks Patrick Dorey for stimulating discussions. The authors thank the participants of the meeting ``Integrability in low-supersymmetry theories'' held in Trani in 2024 and funded by the COST Action CA22113, by INFN and by Salento University.
DP is grateful to the Kavli Institute for
Theoretical Physics in Santa Barbara for the hospitality during the follow-on of the Integrable22 workshop,
where part of this work was carried out.
This work has received funding from the Deutsche Forschungsgemeinschaft (DFG, German Research Foundation) – SFB-Geschäftszeichen 1624 – Projektnummer 506632645 and from the European Union – NextGenerationEU, from the program
STARS@UNIPD, under the project ``Exact-Holography – A
new exact approach to holography: harnessing the power
of string theory, conformal field theory and integrable
models''. The authors also acknowledge support from the PRIN Project n. 2022ABPBEY,
``Understanding quantum field theory through its deformations'' and from the CARIPLO Foundation ``Supporto ai giovani talenti italiani nelle competizioni dell'European Research Council'' grant n. 2022-1886 ``Nuove basi per la teoria delle stringhe''.

\appendix

\section{The S-matrix Jacobian}
\label{app:Jacobian}

 In this short appendix, we show how to convert a 1+1 dimensional amplitude into a 1+1 dimensional S-matrix. 
Consider a process involving the scattering of $n$ particles, which we consider all incoming.  With our convention for the interaction vertices, to pass from a $n$-particle amplitude $M_n$ to a $n$-particle S-matrix $S_n$ we need to introduce a normalization factor $1/\sqrt{4 \pi}$ for each particle. Moreover, we need to introduce the Dirac delta function of the overall energy-momentum conservation:
\begin{equation}
S_n(p_1, \dots, p_n)= \frac{M_n(p_1, \dots, p_n)}{(4 \pi)^{n/2}} \delta ( \sum^n_{i=1} p_i ) \ \delta ( \sum^n_{i=1} E_i ) \,.
\end{equation}
Evaluating the second Dirac delta function on the nonvanishing support of the first Dirac delta function we obtain
\begin{equation}
\begin{split}
\delta ( \sum^n_{i=1} p_i ) \ \delta ( \sum^n_{i=1} E_i )&=\delta ( \sum^n_{i=1} p_i ) \ \delta ( \sqrt{(\sum^n_{i=2} p_i)^2 +m^2_1 }+\sqrt{p_2^2 +m^2_2 }+\sum^n_{i=3} E_i )\\
&=\sum_{i=\text{R, T}}\frac{E^{(i)}_1 E^{(i)}_2}{|E^{(i)}_1 p^{(i)}_2 - E^{(i)}_2 p^{(i)}_1|} \, \delta ( p_1 - p^{(i)}_1 ) \delta(p_2 - p^{(i)}_2) \,,
\end{split}
\end{equation}
where in the last equality we expressed the second Dirac delta function in terms of the momentum $p_2$. For each $i=$ R, T, $p^{(i)}_1$ and $p^{(i)}_2$ are the solutions of $p_1$ and $p_2$ to the energy-momentum conservation constraint and are functions of the remaining momenta $p_3, \, \dots, \, p_n$.  
Since the constraint is of second order, there are two such solutions and we label them by R and T because in two-to-two processes with external particles of the same types these solutions are associated with reflection and transmission. After expressing the energies and momenta in terms of rapidities
\begin{equation}
E_1=\hat{m}_1 \cosh{\theta_1}\,, \qquad p_1=\hat{m}_1 \sinh{\theta_1}, \qquad E_2=\hat{m}_2 \cosh{\theta_2}\,, \qquad p_2=\hat{m}_2 \sinh{\theta_2}
\end{equation}
we obtain
\begin{equation}
|E^{(i)}_1 p^{(i)}_2 - E^{(i)}_2 p^{(i)}_1|= \hat{m}_1 \hat{m}_2 |\sinh{\theta_{12}}|
\end{equation}
and
\begin{equation}
\delta ( p_1 - p^{(i)}_1 )= \frac{\delta ( \theta_1 - \theta^{(i)}_1 )}{E^{(i)}_1}\,, \qquad \delta ( p_2 - p^{(i)}_2 )= \frac{\delta ( \theta_2 - \theta^{(i)}_2 )}{E^{(i)}_2}\,.
\end{equation}
Then, the S-matrix is connected to the amplitude by the following relation
\begin{equation}
S_n(p_1, \dots, p_n)= \frac{M_n(p_1, \dots, p_n)}{(4 \pi)^{n/2}} \frac{1}{\hat{m}_1 \hat{m}_2} \, \sum_{i=\text{R, T}}\frac{1}{|\sinh{\theta_{12}}|} \, \delta ( \theta_1 - \theta^{(i)}_1 ) \delta(\theta_2 - \theta^{(i)}_2)\,.
\end{equation}
For a two-to-two purely elastic process in which the $1^{\text{st}}$ particle is of the same type as the $4^{\text{th}}$ particle and the $2^{\text{nd}}$ particle is of the same type as the $3^{\text{rd}}$ one we get 
\begin{equation}
S_4(p_1, p_2;  p_3, p_4)= \frac{M_4(p_1, p_2 ; p_3,  p_4)}{4 \pi \hat{m}_1 \hat{m}_2 |\sinh{\theta_{12}}|}  \, \delta ( \theta_1 - \theta_4 ) \delta(\theta_2 - \theta_3)\,.
\end{equation}

\section{Planar geometry and on-shell kinematics}
\label{app:geometrical_relations}

In this appendix, we briefly review the conditions for the cancellation of inelastic tree-level amplitudes in integrable theories. We also recall some simple relations on planar geometry necessary for the study of tree-level and one-loop processes.

\subsection{Geometrical relations}
We start by recalling the following property whose proof can be found for example in~\cite{Dorey:2021hub}.
\begin{mytheorem}
\label{theo1}
Given four points $\{ X_0, X_1, X_2 , X_3\}$ in $\mathbb{R}^2$ and setting $X_{ij} = ||X_i - X_j||^2$ it holds that
\begin{equation}
\label{app_eq_dXij}
    \sum_{i\neq j \neq k \neq n} \varepsilon_{ij}(k,n) \Delta_{kni} \Delta_{knj}\ dX_{ij} = 0 \,,
\end{equation}
where $\varepsilon_{ij}(k,n)=1$ if $X_i$ and $X_j$ lie on the same side of the line identified by the vector $X_{k} - X_{n}$, and $\varepsilon_{ij}(k,n)=-1$ if $X_i$ and $X_j$ lie on the opposite sides of this line. $\Delta_{ijk}$ is the area of the triangle having as vertices the points $X_i$, $X_j$ and $X_k$
\end{mytheorem}

\begin{figure}
\begin{center}
\includegraphics*[width=0.8\textwidth]{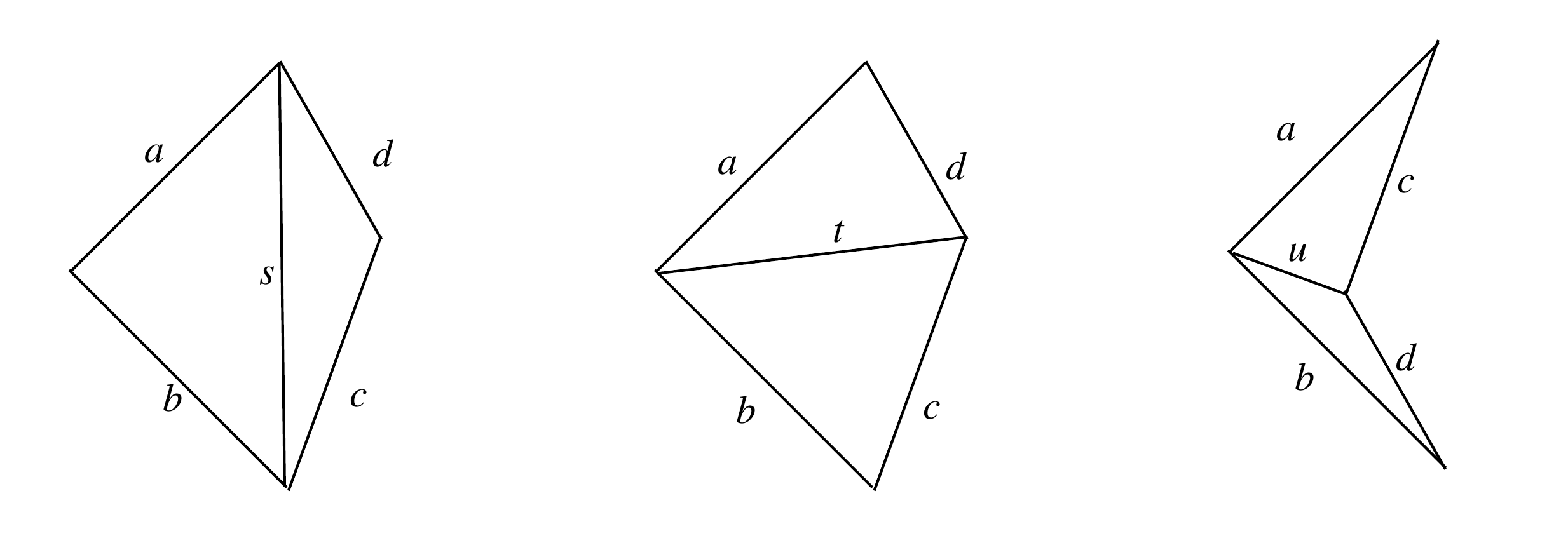} 
\end{center}
\caption{Kinematical configurations associated with the $s$, $t$ and $u$ channels from left to right, respectively.} 
\label{stu_channels}
\end{figure}
Defining the Mandelstam variables $s$, $t$ and $u$ as in~\eqref{eq:s_t_u_Mandelstam}, and applying property \ref{theo1} to the Euclidean kinematical configurations in figure \ref{stu_channels} we obtain the following relations\footnote{Note that differently from the convention in property~\ref{theo1} now the subscript indices correspond to the triangle sides and not to the vertices.}
\begin{multline}
\label{eq:dsdtrelation_appendix}
    \Delta_{sab} \Delta_{scd}\ dt = - \Delta_{tbc} \Delta_{tad}\ ds +\Delta_{tbc} \Delta_{scd}\ d \mu_a^2+  \Delta_{tad} \Delta_{scd}\ d\mu_b^2  + \Delta_{sab} \Delta_{tad}\ d\mu_c^2 \\
    + \Delta_{tbc} \Delta_{sab}\ d\mu_d^2 \,,
\end{multline}
\begin{multline}
\label{eq:dsdurelation_appendix}
    \Delta_{sab} \Delta_{scd} \ du = \Delta_{uac} \Delta_{ubd}\ ds + \Delta_{scd} \Delta_{ubd}\ d\mu_a^2+ \Delta_{scd} \Delta_{uac}\ d\mu_b^2 - \Delta_{sab} \Delta_{ubd}\ d\mu_c^2  \\
    - \Delta_{uac} \Delta_{sab}\ d\mu_d^2 \,,
\end{multline}
\begin{multline}
\label{eq:dtdurelation_appendix}
     \frac{du}{\Delta_{uac} \Delta_{ubd}}+\frac{dt}{\Delta_{tad} \Delta_{tbc}} = \frac{d \mu^2_a}{\Delta_{uac}\Delta_{tad}}+\frac{d \mu^2_b}{\Delta_{tbc}\Delta_{ubd}}-\frac{d \mu^2_c}{\Delta_{uac}\Delta_{tbc}}-\frac{d \mu^2_d}{\Delta_{ubd}\Delta_{tad}}\,,
\end{multline}
where~\eqref{eq:dtdurelation_appendix} can be obtained by combining~\eqref{eq:dsdtrelation_appendix} and~\eqref{eq:dsdurelation_appendix}. 
From these relations, we can immediately read the derivatives in \eqref{eq:t-der1} - \eqref{eq:u-der2}.

\subsection{Tree-level cancellations and flipping rule}
\label{app:tree_canc_and_flipping}

In this section we review the cancellation of singularities in tree-level inelastic processes. Consider the following tree-level amplitude 
\begin{equation}
\label{eq:160215}
    M_{ab\rightarrow cd}^{(0)} = -i \sum_{i\in s} \frac{C^{(3)}_{ab\bar{i}} C^{(3)}_{i\bar{c}\bar{d}}}{s-\mu_i^2} 
    -i \sum_{j\in t} \frac{C^{(3)}_{b\bar{d}\bar{j}} C^{(3)}_{ja\bar{c}}}{t-\mu_j^2}
    -i \sum_{l\in u} \frac{C^{(3)}_{b\bar{c}\bar{l}} C^{(3)}_{la\bar{d}}}{u-\mu_l^2} - i C^{(4)}_{ab\bar{c}\bar{d}} \,,
\end{equation}
associated with the inelastic process~\eqref{eq:ab_to_cd_process}. A necessary condition for this amplitude to vanish is that it contains no singularities for $\theta_{p p'} \in \mathbb{C}$. However, we notice that this amplitude can potentially have simple poles, due to the propagation of on-shell bound states in the $s$, $t$ and $u$ channels. This corresponds to the case in which one of the three diagonals in figure~\ref{stu_channels} becomes equal to the mass of a propagating particle.

Following~\cite{Dorey:2021hub}, let us see how the poles of~\eqref{eq:160215} cancel. For this to happen it is necessary that whenever a pole exists in a certain Feynman diagram (due to the propagation of an on-shell bound state in some channel) then there is at least another diagram becoming singular at the same value of the external kinematics and the total residue given by the sum of the two diagrams vanishes.
This is the so-called \textit{flipping rule} discussed in~\cite{Braden:1990wx} and used to study this type of tree-level cancellations in~\cite{Dorey:2021hub}.

To construct these singular diagrams we note that at the poles the rapidities become purely imaginary and one can write each momentum involved in the scattering as a vector on the complex plane, whose absolute value is the mass of the scattered particle. Due to momentum conservation, these vectors form a closed polygon and each class of Feynman diagrams, with particles propagating in the $s$, $t$ and $u$ channels, corresponds to dividing the polygon into two sides with a diagonal. For instance, for the $s$, $t$ and $u$ channel in the tree-level amplitude \eqref{eq:160215} we associate the three quadrilaterals in figure~\ref{stu_channels}. The squared of the diagonals are the Mandelstam variables and each of them (from left to right in figure~\ref{stu_channels}) is $\sqrt{s}$, $\sqrt{t}$ and $\sqrt{u}$.
Here we assume $\sqrt{t}$ to be the diagonal associated with the convex shape and $\sqrt{u}$ to be the diagonal associated with the concave shape. 

Now when we move to the pole position $s=m_i^2$, these diagrams become on-shell and the diagonals become the masses of propagating particles. This is where the flipping rule enters: if we are at $s=m_i^2$  then either $\sqrt{t}$ or $\sqrt{u}$ (or both) becomes a physical mass and the associated singular Feynman diagram cancels with the $s$ channel pole.
Let us assume that simultaneous singularities appear in both the $t$ and $u$ channels, with propagating particles $j$ and $l$ respectively.
Then for $s=m_i^2$ we have $t=m_j^2$ and $u=m_l^2$.
Parameterising the $3$-point couplings as
\begin{equation}
\label{eq:417068}
     C^{(3)}_{abc}=f_{abc} \, \Delta_{abc}\,.
\end{equation}
with  $\Delta_{abc}$ being the area of the triangle having sides $m_a$, $m_b$ and $m_c$, 
then the condition for the cancellation of the three singularities is~\cite{Dorey:2021hub} 
\begin{equation}
\label{eq:94754}
     f_{ab\bar{i}} f_{i\bar{c}\bar{d}} 
    - f_{a\bar{d}\bar{j}} f_{jb\bar{c}} + f_{a\bar{c}\bar{l}} f_{lb\bar{d}} = 0.
\end{equation}
The formula above generalizes also to the cases where the singularities only appear in two channels (which can be $s/t$, $s/u$ or $t/u$): in such a case we omit the term associated with the channel that is not singular. Degeneracies can also be taken into account introducing sums over particles with the same mass. 
Using these geometrical criteria we can see which diagrams are simultaneously singular and should cancel to avoid inelastic processes at the tree level.

\section{Mass shifts in nonsimply-laced theories}
\label{app:mass-shifts}

In this appendix, we collect the mass shifts of nonsimply-laced affine Toda theories and 
use them to compute the mass ratios to one loop order in perturbation theory. 
In all cases, the perturbative results match the ratios of the physical masses proposed in~\cite{Delius:1991kt,Corrigan:1993xh} to one loop (all checks are analogue to the one performed in section \ref{subsec_one_loop_Smat} for $g_2^{(1)}$ and $d_4^{(3)}$).
We begin by recalling the general expression for one-loop mass corrections and then compute these corrections for all untwisted and twisted nonsimply-laced affine Toda models. 
While these perturbative corrections to the masses are well known for a while (see, e.g., \cite{Delius:1991ie,Grisaru:1990sv}), we report them here for completeness.

\subsection{Mass corrections from S-matrices}

Following~\cite{Polvara:2023vnx} we write the one-loop mass shifts in terms of tree-level on-shell amplitudes as
\begin{equation}
\label{eq:mass-corr-1}
\delta m_a^2 = -\frac{i}{8 \pi} \sum^r_{b=1} \int^{+\infty}_{-\infty} d \theta \left( M^{(0)}_{ab}(\theta) - M^{(0)}_{ab}(\infty)\right) \,.
\end{equation}
One can rewrite this as a contour integral using crossing symmetry, which (at the tree level) for theories with only real particles (as the cases treated here) takes the form
\begin{equation}
M^{(0)}_{ab}(\theta)=M^{(0)}_{ab}(\theta+ i \pi).
\end{equation}
Therefore, by introducing an auxiliary parameter $\beta$ and using crossing, we can write
\begin{equation}
\int^{+\infty}_{-\infty} d \theta\ e^{i \beta \theta} \left( M^{(0)}_{ab}(\theta) - M^{(0)}_{ab}(\infty)\right) = \frac{1}{1 - e^{- \pi \beta}} \oint_{\Gamma} d \theta\ e^{i \beta \theta} M^{(0)}_{ac}(\theta) \,,
\end{equation}
where $\Gamma$ is a rectangle of infinite horizontal size and vertical side going from $0$ to $i \pi$. This contour is closed in the counterclockwise direction with the lower horizontal side corresponding to the real axis and the upper horizontal side corresponding to the line with a constant imaginary part equal to $i \pi$. In the limit $\beta \to 0$ we obtain
\begin{equation}
\label{eq:mass_correction_as_closed_integral1}
\delta m_a^2 = \frac{1}{8 \pi^2} \sum_{b=1}^r \oint_{\Gamma} d \theta \, \theta \, M^{(0)}_{ab}(\theta) \,.
\end{equation}
In terms of the S-matrix the mass corrections are
\begin{equation}
\label{eq:mass_correction_as_closed_integral2}
\delta m_a^2 = \frac{m_a}{2 \pi^2} \sum^r_{b=1} m_b \oint_{\Gamma} d \theta \, \theta \, \sinh\theta\ S^{(0)}_{ab}(\theta) \, .
\end{equation}
The tree-level S-matrices used to compute the mass shifts in nonsimply-laced models are obtained by expanding the exact S-matrices proposed  \cite{Delius:1991kt,Corrigan:1993xh} to the leading order in the coupling $\g$. Since these match at tree level with the universal expressions of \cite{Dorey:2021hub} there is no problem in using them. A small caveat is that for some models one has to rescale $g$ in the S-matrices of \cite{Delius:1991kt,Corrigan:1993xh} to agree with the convention that the length of the long roots is always $\sqrt{2}$ and the factors $\sigma_{abc}$ can only take the values $\pm 1$, $\pm1/\sqrt{2}$ and $\pm 2/\sqrt{3}$ (see property C.4 in~\cite{Dorey:2021hub}).

\subsection{Untwisted theories}
\label{sec:untwisted-theories}

Nonsimply-laced affine Toda models are obtained from simply-laced theories through a procedure called folding, as explained in \cite{Braden:1989bu}. The main idea is that a given symmetry $p$ of a simply-laced Dynkin diagram, which acts on the simple roots by $p(\alpha)$, induces an action on the fields of the related affine Toda model by $p(\phi)$. 
By projecting the fields to the space invariant under $p$ we obtain the nonsimply-laced theories.
The symmetries used for folding are the ones satisfying $\alpha\cdot p(\alpha)=0$ and are called direct reductions. 
The set of untwisted nonsimply-laced theories is obtained when $p$ is a symmetry of the unextended Dynkin diagram. In this case, the resulting Dynkin diagram after the folding
turns out to be the extension of a non-simply laced Dynkin diagram. Among nonsimply-laced theories, there are four untwisted models: $b^{(1)}_n$, $c^{(1)}_n$, $g^{(1)}_2$ and $f^{(1)}_4$. This type of folding removes degeneracies of the simply-laced models and the mass spectrum is a subsector of the parent theory; however, the cubic couplings are different from the ones of the starting model. Below we provide the mass shifts for each untwisted theory. 

\paragraph{The $b^{(1)}_n$ class.} 
These models are obtained from the folding of the unextended Dynkin diagram of the $d_{n+1}^{(1)}$ simply-laced theories. They contain $n$ particles from the starting model with classical masses
\begin{equation}
    \begin{split}
        & m_a^2 = 4 \mu^2 \sin^2 \left( \frac{\pi a}{h} \right) \ \ \textrm{with} \ \ a=1,\cdots, n-1, \\
        & m_n^2 = \mu^2.
    \end{split}
\end{equation}
Above $h=2n$ is the Coxeter number of the starting $d_{n+1}^{(1)}$ theory. The cubic couplings follow the area rule~\eqref{eq:coupling_area_rule2} with
\begin{equation}
    \begin{split}
        & \sigma_{abc}=
        \begin{cases}
        & -1 \ \  \textrm{if} \ a+b+c=h , \\
        & +1 \ \ \textrm{if} \ a \pm b \pm c=0 ,\\
        & 0 \ \ \quad  \textrm{otherwise} ,
        \end{cases} \\
        & \sigma_{nna} = 1, \\
        & \sigma_{nnn} = \sigma_{nab} = 0.
    \end{split}
\end{equation}
The mass shifts are found following the procedure previously described and do not scale equally; indeed we have
\begin{equation}
\label{eq:170731}
    \begin{split}
        & \delta m_a^2 = \frac{g^2 \mu ^2}{h^2} \left(h \cot \left(\frac{\pi
   }{h}\right) \sin ^2\left(\frac{\pi 
   a}{h}\right)-a \sin \left(\frac{2 \pi 
   a}{h}\right)\right) \ \ \textrm{for} \ \ a=1,\cdots, n-1, \\
        & \delta m_n^2 = \frac{g^2 \mu ^2}{4 h} \cot \left(\frac{\pi
   }{h}\right).
    \end{split}
\end{equation}
Using these mass shifts, we find the following expansions for the ratios of the physical masses
\begin{equation}
    \frac{\hat{m}_a^2}{\hat{m}_n^2} = 4 \sin^2 \left( \frac{\pi a}{h} \right)+\frac{a g^2}{4 n^2}\sin \left(\frac{\pi  a}{n}\right)+O\left(g^3\right).
\end{equation}

\paragraph{The $c^{(1)}_n$ class.} 
These theories are obtained by folding the unextended Dynkin diagrams of $a_{2n-1}^{(1)}$ models and classically have the following spectrum
\begin{equation}
    m_a^2 = 4 \mu^2 \sin^2 \left(\frac{\pi a}{h}\right), \ \ \textrm{with} \ \ a=1,\cdots, n-1,
\end{equation}
with $h=2n$. The coefficients of the cubic couplings are
\begin{equation}
    \begin{split}
        & \sigma_{abn} = -1, \ \ \textrm{if} \ \ a+b+n = 0 \mod 2n, \\
        & \sigma_{abc} = \frac{1}{\sqrt{2}} \ \ \ \textrm{if} \ \ a\pm b\pm c = 0 ,\\
        &\sigma_{ann}=\sigma_{nnn}=0
    \end{split}
\end{equation}
and the corrections to the masses are
\begin{equation}
    \delta m_a^2 = \frac{\mu^2 g^2}{2 h^2} \left( 2 a \sin\left(\frac{2\pi a}{h}\right) + h \cot\left(\frac{\pi}{h}\right) \sin^2 \left(\frac{\pi a}{h}\right) \right),
\end{equation}
The expansions of the physical mass ratios are
\begin{equation}
    \frac{\hat{m}_a^2}{\hat{m}_1^2} = \frac{\sin^2 (\pi a/h)}{\sin^2(\pi/h)}+\frac{g^2 }{h^2}\frac{\sin^2(\pi a/h)}{\sin^2(\pi/h)}\left(\cot \left(\frac{\pi }{h}\right)-a \cot \left(\frac{\pi  a}{h}\right)\right)+O\left(g^3\right).
\end{equation}

\paragraph{The $g^{(1)}_2$ model.}

From the folding of the unextended Dynkin diagram of the $d_4^{(1)}$ theory one derives this model which has only two particles with classical masses
\begin{equation}
    \begin{split}
        & m_1^2 = 2 \mu^2, \\
        & m_2^2 = 6 \mu^2.
    \end{split}
\end{equation}
and cubic couplings obtained from
\begin{equation}
\label{eq:g2-couplings}
    \begin{split}
        & \sigma_{222}= -\sigma_{112}= 1, \\
        & \sigma_{111} = - \frac{2}{\sqrt{3}}.
    \end{split}
\end{equation}
through the area rule.
The mass shifts are easily computed in this case and we find that
\begin{equation}
\label{eq:312245}
    \begin{split}
        & \delta m_1^2 = \frac{7g^2 \mu^2}{36\sqrt{3}}, \\
        & \delta m_2^2 = \frac{5g^2 \mu^2}{12\sqrt{3}} .
    \end{split}
\end{equation}
The physical mass ratio is then found to be
\begin{equation}
    \frac{\hat{m}_2^2}{\hat{m}_1^2} = 3+\frac{g^2}{4 \sqrt{3}}+O\left(g^3\right).
\end{equation}

\paragraph{The $f^{(1)}_4$ model.}
This model is obtained by folding the unextended Dynkin diagram of the simply-laced $e_6^{(1)}$ theory. It has the following set of classical masses
\begin{equation}
    \begin{split}
        m_1^2 & = (3-\sqrt{3}) \mu^2 , \\
        m_2^2 & = 2(3-\sqrt{3}) \mu^2 , \\
        m_3^2 & = (3+\sqrt{3})\mu^2 , \\
        m_4^2 & = 2(3+\sqrt{3})\mu^2 .
    \end{split}
\end{equation}
The list of nonzero cubic couplings is long and we give them in table \ref{table:f4 coupling}. 
\begin{table}[t!]
\centering
\begin{tabular}{c|*{4}{c|}}
 \M{} & \M{$_1$} & \M{$_2$} & \M{$_3$} & \M{$_4$} \\ \cline{2-5}
    %\hline
  \MI{$_1$} & $1^*$ $2$ $3^*$ & $1$ $3$ & $1^*$ $2$ $3^*$ $4$ & $3$     \\ \cline{2-5}
  \M{} & \MI{$_2$} & $2$ $4$ & $1$ & $2$ $4^{-}$    \\ \cline{3-5}
  \M{} & \M{} &  \MI{$_3$} & $1^*$ $3^*$ $4^{-}$ & $1$ $3^{-}$ \\ \cline{4-5}
  \M{} & \M{} & \M{} &  \MI{$_4$} & $2^{-}$ $4$ \\ \cline{5-5}
\end{tabular}
\caption{Non-zero couplings of $f_4^{(1)}$ model, taken from~\cite{Dorey-thesis}. In each cell, there are the particles that couple with the row and column indices. In the cases without exponents we mean $\sigma_{abc}=1$ while the superscript indices `$*$' and `$-$' correspond to $\sigma_{abc}=1/\sqrt{2}$ and $\sigma_{abc}=-1$, respectively. For example, we can see from the third row and third column that $\sigma_{331}=\sigma_{333}=1/\sqrt{2}$ and $\sigma_{334}=-1$.}
\label{table:f4 coupling}
\end{table}
The mass shifts were found to be
\begin{equation}
\label{eq:84921}
    \begin{split}
        \delta m_1^2 & = \frac{6+\sqrt{3}}{96} \mu^2 g^2 , \\
        \delta m_2^2 & = \frac{3+2\sqrt{3}}{48} \mu^2 g^2 , \\
        \delta m_3^2 & = \frac{4+3\sqrt{3}}{32} \mu^2 g^2 , \\
        \delta m_4^2 & = \frac{5+2\sqrt{3}}{16} \mu^2 g^2 .
    \end{split}
\end{equation}
and the ratios of the physical masses are given by
\begin{equation}
    \begin{split}
        & \frac{\hat{m}_2^2}{\hat{m}_1^2} = 2+\frac{g^2}{24}+O\left(g^3\right), \\
        & \frac{\hat{m}_3^2}{\hat{m}_1^2} = 2+\sqrt{3}+\frac{g^2}{48}+O\left(g^3\right), \\
        & \frac{\hat{m}_4^2}{\hat{m}_1^2} = 2 \left(2+\sqrt{3}\right)+ \left(1+\sqrt{3}\right)\frac{g^2}{24} +O\left(g^3\right).
    \end{split}
\end{equation}

\subsection{Twisted theories}
\label{sec:twisted-theories}

The other class of nonsimply-laced affine Toda models 
is composed of twisted theories. In this case, the map $p$ used for the reduction is a symmetry of an affine simply-laced Dynkin diagram and it turns out that the resulting theory is a truncation of the starting simply-laced model to a closed subsystem of particles. More precisely, the resulting twisted theory has a spectrum contained in the parent theory with couplings and tree-level S-matrices being a subset of the original ones~\cite{Braden:1989bu}. There are five models in this class: $a^{(2)}_{2n}$, $d^{(2)}_{n+1}$, $a^{(2)}_{2n-1}$, $d^{(3)}_4$ and $e^{(2)}_6$.  

\paragraph{The $a^{(2)}_{2n}$ class.} These models are obtained from the folding of $d^{(1)}_{2n+2}$ which has $2n+2$ particles. Apart from the $(2n+2)^{\text{th}}$ particle (which is removed) the folding preserves the even particles in the spectrum; thus $a^{(2)}_{2n}$ has $n$ particles with classical masses 
\begin{equation}
\label{eq:mass_spectrum_A2n2}
m^2_a=8 \mu^2 \sin^2 \left( \frac{a \pi}{h} \right) \,, \quad a=2,\, 4\,, \dots, 2n\,,
\end{equation}
with $h=4n+2$ being the Coxeter number of the starting simply-laced theory.
As aforementioned, the couplings are just a subset of the $d^{(1)}_{2n+2}$ cubic couplings and are given by
\begin{equation}
\label{eq:couplings_D2nplus2}
\sigma_{a b c}=
\begin{cases}
&-1 \  \ \  \text{if} \ a+b+c=h\,,\\
&+1 \  \ \ \text{if} \ a \pm b \pm c=0\,,\\
&0 \ \ \quad \text{otherwise} \,.
\end{cases}
\end{equation}
The mass shifts are
\begin{equation}
    \delta m_a^2 = \frac{g^2}{4 h} \cot\left(\frac{2\pi}{h}\right) m_a^2 \,.
\end{equation}
From the equation above, we see that all the masses scale equally in this class and (as it happens in simply-laced models) the mass ratios do not receive quantum corrections.

\paragraph{The $d^{(2)}_{n+1}$ class.}

These theories are obtained by folding the extended Dynkin diagrams of $d^{(1)}_{n+2}$ simply-laced models. From the original spectrum (composed of $n+2$ particles) we keep $n$ particles, with classical masses
\begin{equation}
m^2_a= 8 \mu^2 \sin^2 \left( \frac{\pi a}{h}  \right) \, \quad a=1, \dots, n \,.
\end{equation}
Here $h=2n+2$ is the Coxeter number of $d^{(1)}_{n+2}$. The cubic couplings are 
\begin{equation}
\sigma_{a b c}=
\begin{cases}
&-1 \ \ \  \text{if} \ a+b+c=h\,,\\
&+1 \ \ \ \text{if} \ a \pm b \pm c=0\,,\\
&0 \ \ \quad \text{otherwise}
\end{cases}
\end{equation}
and are a subset of the couplings of $d^{(1)}_{n+2}$. The mass shifts in this case are
\begin{equation}
\label{eq:mass_shifts_Dnplus1_2}
    \delta m_a^2 = \frac{2\mu^2 g^2}{h}\sin^2\left(\frac{\pi a}{h}\right) \cot\left(\frac{\pi }{h}\right) - \frac{4\mu^2 g^2}{h^2}\ a \sin\left(\frac{2\pi a}{h}\right)
\end{equation}
and do not scale uniformly. The physical mass ratios have the following expansions
\begin{equation}
    \frac{\hat{m}_a^2}{\hat{m}_1^2} = \frac{\sin^2 (\pi a/h)}{\sin^2(\pi/h)}-\frac{g^2}{h^2} \frac{\sin^2 (\pi a/h)}{\sin^2(\pi/h)} \left(\cot
   \left(\frac{\pi }{h}\right)-a \cot \left(\frac{\pi  a}{h}\right)\right) +O\left(g^3\right).
\end{equation}

\paragraph{The $a^{(2)}_{2n-1}$ class.}
These theories are obtained by folding the extended Dynkin diagrams of the $d^{(1)}_{2n}$ models. We start with $4n-2$ particles (before the folding) and we end up with $n$ particles with indexes $\{2,4,\cdots,2n-2,s\}$, with $s$ being the lightest particle. The classical mass spectrum of this twisted theory is
\begin{equation}
\begin{split}
    & m_a^2 = 8\mu^2\sin^2 \left(\frac{\pi a}{h}\right)\ \ \textrm{for} \ \ a\in\{2,4,\cdots,2n-2\}, \\
    & m_s^2 = 2\mu^2 ,
\end{split}
\end{equation}
with the Coxeter number given by $h=4n-2$. The couplings are 
\begin{equation}
\begin{split}
& \sigma_{a b c}=
\begin{cases}
&-1 \ \ \  \text{if} \ a+b+c=h\,,\\
&1 \ \ \ \text{if} \ a \pm b \pm c=0\,,\\
&0  \ \ \ \text{otherwise} \,,
\end{cases} \\
& \sigma_{ssa} = 1
\end{split}
\end{equation}
and the mass shifts were found to be
\begin{equation}
\begin{split}
    & \delta m_a^2 = \frac{2 \mu^2 g^2}{h} \cot\left( \frac{2\pi}{h} \right) \sin^2\left( \frac{\pi a}{h} \right) + \frac{2 \mu^2 g^2}{h^2}\ a \sin\left( \frac{2\pi a}{h} \right), \\
   & \delta m_s^2 = \frac{\mu^2 g^2}{2h} \left( \csc\left( \frac{2\pi}{h} \right) - \tan\left( \frac{\pi}{h} \right) \right) \,.
\end{split}
\end{equation}
Then the physical mass ratios are
\begin{equation}
    \frac{\hat{m}_a^2}{\hat{m}_s^2} = 4 \sin ^2\left(\frac{\pi  a}{h}\right)-\frac{a g^2}{h^2}\sin \left(\frac{2
   \pi  a}{h}\right)+O\left(g^3\right).
\end{equation}

\paragraph{The $d^{(3)}_4$ model.}

This theory is obtained by folding the extended Dynkin diagram of $e^{(1)}_6$. Among the  $6$ particles of the starting model, the folding preserves only the particles $2$ and $4$, which have non-degenerate masses. The classical masses of these particles are
\begin{equation}
\begin{split}
    & m_2^2 = 2(3-\sqrt{3})\mu^2 , \\
    & m_4^2 = 2(3+\sqrt{3})\mu^2 .
\end{split}
\end{equation}
The non-zero cubic couplings in this case are
\begin{equation}
    \sigma_{222} = \sigma_{224} = -\sigma_{244} = \sigma_{444} = - 1, 
\end{equation}
and the mass shifts turn out to be
\begin{equation}
\begin{split}
    & \delta m_2^2 = \frac{\mu^2 g^2}{8\sqrt{3}}, \\
    & \delta m_4^2 = \frac{6+\sqrt{3}}{24} \mu^2 g^2 .
\end{split}
\end{equation}
The physical mass ratio is simply
\begin{equation}
    \frac{\hat{m}_4^2}{\hat{m}_2^2} = 2+\sqrt{3}-\frac{g^2}{48}+O\left(g^3\right).
\end{equation}

\paragraph{The $e^{(2)}_6$ model.}

The last twisted model is obtained by folding the extended Dynkin diagram of the $e^{(1)}_{7}$ simply-laced theory; in the folding, the spectrum reduces from seven to four particles, labelled by $\{2,4,5,7\}$. The classical masses are
\begin{equation}
\begin{split}
    & m_2^2 = 8\sqrt{3} \sin\left(\frac{\pi}{18}\right) \sin\left(\frac{2\pi}{9}\right) \mu^2 ,\\
    & m_4^2 = 8\sqrt{3} \sin\left(\frac{5\pi}{18}\right) \sin\left(\frac{\pi}{9}\right) \mu^2 , \\
    & m_5^2 = 6 \mu^2, \\
    & m_7^2 = 8\sqrt{3} \sin\left(\frac{7\pi}{18}\right) \sin\left(\frac{4\pi}{9}\right) \mu^2.
\end{split}
\end{equation}
We list the cubic couplings in table~\ref{table:e62 coupling}. The exact expressions for the mass shifts in this case are convoluted, but we nonetheless show them below for completeness:
\begin{table}[t!]
\centering
\begin{tabular}{c|*{4}{c|}}
 \M{} & \M{$_2$} & \M{$_4$} & \M{$_5$} & \M{$_7$} \\ \cline{2-5}
    %\hline
  \MI{$_2$} & $2^-$ $4$ $5^-$ & $2$ $5$ & $2^-$ $4$ $7^-$ & $5^-$ $7$ \\ \cline{2-5}
  \M{} & \MI{$_4$} & $4$ $5$ $7^-$ & $2$ $4$ $7^-$ & $4$ $5^-$  \\ \cline{3-5}
  \M{} & \M{} &  \MI{$_5$} & $5$ & $2^-$ $4^-$ $7^-$ \\ \cline{4-5}
  \M{} & \M{} & \M{} &  \MI{$_7$} & $2$ $5^-$ $7$ \\ \cline{5-5}
\end{tabular}
\caption{Non-zero couplings of $e_6^{(2)}$ model, taken from~\cite{Dorey-thesis}. Here we follow the same notation as in table \ref{table:f4 coupling}; all couplings have the same magnitude $|\sigma_{abc}|=1$ with minus signs indicated by the superscript indices.}
\label{table:e62 coupling}
\end{table}
{\small\begin{equation}
    \begin{split}
        \delta m_2^2 & = \frac{g^2 \mu^2}{324}  \biggl(-\sqrt{3} \left(9+4 \sin
   \left(\frac{\pi }{18}\right)\right)-42 \sin \left(\frac{\pi
   }{9}\right)-18 \sin \left(\frac{2 \pi }{9}\right)+38 \sqrt{3}
   \cos \left(\frac{\pi }{9}\right)\\
   & +4\ 3^{3/4} \sqrt{\sin
   \left(\frac{\pi }{18}\right) \sin \left(\frac{2 \pi
   }{9}\right)} \left(-3+8 \sin \left(\frac{\pi }{18}\right)+4
   \cos \left(\frac{2 \pi }{9}\right)\right)\biggr), \\
   %\approx 0.0726834\ \mu^2 g^2 , \\
   %%%%%%%%%%%%%%%%
    \delta m_4^2 & = -\frac{g^2 \mu^2}{648} \csc \left(\frac{\pi }{18}\right) \sec
   \left(\frac{\pi }{9}\right) \biggl(-9 \sin \left(\frac{\pi
   }{9}\right)-12 \sqrt{3} \cos \left(\frac{\pi }{9}\right)+6
   \sqrt{3} \cos \left(\frac{2 \pi }{9}\right)\\
   & + 4\ 3^{3/4}
   \left(-3+7 \sin \left(\frac{\pi }{18}\right)-2 \cos
   \left(\frac{\pi }{9}\right)+3 \cos \left(\frac{2 \pi
   }{9}\right)\right) \sqrt{\sin \left(\frac{\pi }{9}\right) \cos
   \left(\frac{2 \pi }{9}\right)}\biggr), \\
   % & \approx 0.198816\ \mu^2 g^2,
    \delta m_5^2 & = \frac{g^2 \mu^2}{108 \sqrt{6}} \biggl(9 \sqrt{2}+2 \sqrt[4]{3} \sqrt{2 \cos
   \left(\frac{\pi }{18}\right) \cos \left(\frac{\pi }{9}\right)}
   \left(6+16 \sin \left(\frac{\pi }{18}\right)+4 \cos
   \left(\frac{\pi }{9}\right)\right)\\
   & +2 \sqrt[4]{3} \left(6-4 \cos
   \left(\frac{\pi }{9}\right)+8 \cos \left(\frac{2 \pi
   }{9}\right)\right) \sqrt{2 \sin \left(\frac{\pi }{9}\right)
   \cos \left(\frac{2 \pi }{9}\right)}\\
   & +2 \sqrt[4]{3} \sqrt{2 \sin
   \left(\frac{\pi }{18}\right) \sin \left(\frac{2 \pi
   }{9}\right)} \left(-6+16 \sin \left(\frac{\pi }{18}\right)+8
   \cos \left(\frac{2 \pi }{9}\right)\right)\biggr), \\
   \delta m_7^2 & = \frac{g^2 \mu^2}{324}  \biggl(27 \sqrt{3}+8 \sqrt{3} \sin
   \left(\frac{\pi }{18}\right)+12 \sin \left(\frac{\pi
   }{9}\right)+48 \cos \left(\frac{\pi }{18}\right)+34 \sqrt{3}
   \cos \left(\frac{\pi }{9}\right)\\
   & +2 \sqrt{3} \cos \left(\frac{2
   \pi }{9}\right)+4\ 3^{3/4} \sqrt{\cos \left(\frac{\pi
   }{18}\right) \cos \left(\frac{\pi }{9}\right)} \left(3+8 \sin
   \left(\frac{\pi }{18}\right)+2 \cos \left(\frac{\pi
   }{9}\right)\right)\biggr).
    \end{split}
\end{equation}} 
It turns out that in the computation of the physical mass ratios most of the terms above cancel out and we find the following simpler expressions
\begin{equation}
    \begin{split}
        \frac{\hat{m}_4^2}{\hat{m}_2^2} & = \frac{1}{16} \csc ^2\left(\frac{\pi }{18}\right) \sec ^2\left(\frac{\pi }{9}\right) \\
  & + \frac{11-4 \sin \left(\frac{\pi }{18}\right)+6 \sqrt{3} \sin \left(\frac{\pi }{9}\right)-6
   \sqrt{3} \sin \left(\frac{2 \pi }{9}\right)-6 \sqrt{3} \cos \left(\frac{\pi }{18}\right)+4
   \cos \left(\frac{\pi }{9}\right)}{108 \left(9 \sqrt{3}+14 \sin \left(\frac{\pi }{9}\right)-20
   \sin \left(\frac{2 \pi }{9}\right)-8 \cos \left(\frac{\pi }{18}\right)\right)} g^2 + O(g^3), \\
        \frac{\hat{m}_5^2}{\hat{m}_2^2} & =  2+2 \cos \left(\frac{\pi }{9}\right)-\frac{g^2}{54} \sin \left(\frac{\pi
   }{9}\right)+O\left(g^3\right), \\
       \frac{\hat{m}_7^2}{\hat{m}_2^2} &  = \frac{1}{4} \csc ^2\left(\frac{\pi }{18}\right)-\frac{g^2}{432} \left(\left(\cot
   \left(\frac{\pi }{18}\right)-3 \sqrt{3}\right) \csc ^2\left(\frac{\pi
   }{18}\right)\right)+O\left(g^3\right) .
    \end{split}
\end{equation}

\bibliographystyle{JHEP}
\bibliography{refs}

\end{document}